\documentclass[a4paper,fleqn,usenatbib]{mnras}

\usepackage{newtxtext,newtxmath}

\usepackage[T1]{fontenc}
\usepackage{ae,aecompl}

\usepackage{graphicx}	
\usepackage{amsmath}	
\usepackage{amssymb}	
\usepackage{afterpage}
\usepackage{verbatim}

\title[From star particles to stars]{Colour-magnitude diagram in simulations of galaxy formation}

\author[M. Valentini et al.]
{Milena Valentini$^{1,2}$\thanks{E-mail: milena.valentini@sissa.it}, 
Alessandro Bressan$^{1,3}$\thanks{E-mail: sbressan@sissa.it},
Stefano Borgani$^{2,4,5}$\thanks{E-mail: borgani@oats.inaf.it},
\newauthor
Giuseppe Murante$^{2}$,
L\'eo Girardi$^{3}$,
and Luca Tornatore$^{2}$
\\ ~ \\
\footnotesize 
$^{1}$ SISSA - International School for Advanced Studies, via Bonomea 265, I-34136 Trieste, Italy\\
$^{2}$ INAF - Osservatorio Astronomico di Trieste, via Tiepolo 11, I-34131 Trieste, Italy\\
$^{3}$ INAF - Osservatorio Astronomico di Padova, Vicolo dell'Osservatorio 5, I-35122 Padova, Italy\\
$^{4}$ Astronomy Unit, Department of Physics, University of Trieste, via Tiepolo 11, I-34131 Trieste, Italy\\
$^{5}$ INFN - National Institute for Nuclear Physics, Via Valerio 2, I-34127 Trieste, Italy\\
}

\date{Accepted 2018 July 12. Received 2018 July 12; in original form 2018 April 21}

\pubyear{2018}

\begin{document}
\label{firstpage}
\pagerange{\pageref{firstpage}--\pageref{lastpage}}
\maketitle

\begin{abstract}

\noindent
State-of-the-art cosmological hydrodynamical simulations have star particles with typical mass 
between $\sim$$10^8$ and $\sim$$10^3$ M$_{\odot}$ according to resolution, and treat them as 
simple stellar populations. On the other hand, observations in nearby galaxies 
resolve individual stars and provide us with single star properties. 
An accurate and fair comparison between predictions from simulations and observations is a crucial task.
We introduce a novel approach to consistently populate star particles with stars.  
We developed a technique to generate a theoretical catalogue of mock stars 
whose characteristics are derived from the properties of parent star particles from 
a cosmological simulation. Also, a library of stellar evolutionary tracks and synthetic spectra 
is used to mimic the photometric properties of mock stars. 
The aim of this tool is to produce a database of synthetic stars from the properties 
of parent star particles in simulations: such a database represents the observable stellar content 
of simulated galaxies and allows a comparison as accurate as possible with observations of 
resolved stellar populations. 
With this innovative approach we are able to provide a colour-magnitude diagram 
from a cosmological hydrodynamical simulation.
This method is flexible and can be tailored to fit output of different codes used for cosmological simulations.
Also, it is of paramount importance with ongoing survey data releases 
(e.g. GAIA and surveys of resolved stellar populations), and will be useful to predict properties of stars 
with peculiar chemical features and to compare predictions from hydrodynamical models with data of 
different tracers of stellar populations.
\end{abstract}




\section{Introduction} 
\label{sec:introduction}
Stars generally have a diverse and complex metallicity and chemical content, this indicating 
the variety of living environments and stellar evolution paths.
While having a main active role in the process of chemical enrichment of the surrounding 
interstellar medium (ISM) of a galaxy, stars also passively contribute to track galaxy evolution, as their 
chemical composition records vital information on the past events of the galactic ecosystem.
The number of long-lived stars and the composition of their stellar atmospheres in terms of different 
metal abundances 
are indeed key tracers of the galaxy star formation history, of the past history of feedback, of the accretion of 
pristine gas from the large-scale environment, and of the timing with which all these processes occurred 
across cosmic time \citep[e.g.][]{Wilson1992, Matteucci2012}. 

Star formation and stellar feedback regulate how subsequent generations of stars pollute and enrich the ISM 
in heavy metals, and galactic outflows and AGN (active galactic nucleus) feedback are the main drivers for the way 
in which metals circulate and are distributed in the ISM and circumgalactic medium (CGM). All these phenomena 
have to be consistently accounted for in a successful model that aims at describing and explaining the framework 
in which the chemical enrichment process fits.
To properly address the study of the chemical enrichment process, several models of galactic chemical evolution 
have been proposed, and the chemical enrichment process has been included in both semi-analytical models of 
galaxy formation and cosmological hydrodynamical simulations of cosmic structure formation 
\citep[see, e.g., the reviews by][and references therein, for chemodynamical and semi-analytical models of galactic 
chemical evolution, and for chemical evolution modelling in cosmological simulations]{Gibson2003, Borgani2008}.

State-of-the-art models of galactic chemical evolution are the most sophisticated tools to investigate the 
evolution of different metal abundances in our Galaxy. Despite being very accurate and complex enough to 
account for processes such as radial gas flows 
\citep{Lacey1985, Portinari2000, Schonrich2009, Spitoni2011} and for different formation and evolutionary 
paths for galactic stellar components, namely halo, thick and thin disc 
\citep{Chiappini1997, Portinari1998, Romano2010},   
models of galactic chemical evolution do not capture the full complexity of galaxy evolution in a cosmological 
context. On the other hand, cosmological hydrodynamical simulations do capture this spatial and temporal 
complexity. They indeed follow the cosmological accretion of gas from the large-scale structure and 
the infall of gas into the innermost regions of forming structures. As this gas cools, it fuels star formation, 
that finally results in a feedback both in energy and in metals. All these processes can be self-consistently 
implemented in cosmological simulations, where chemical evolution is one natural outcome of galaxy evolution.

The distribution of metals in the ISM and in the CGM are crucial features of hydrodynamical simulations 
\citep[e.g.][]{Oppenheimer2017, Torrey2017}.
Albeit less frequently investigated, the comparison with the observed stellar metallicity distribution poses a strong 
constraint for simulations of galaxy formation (see below), and the metal content of stars predicted by models 
is crucial to interpret a wide range of observations. 
Furthermore, a great effort has been recently devoted to simulate the actual star counts in several 
regions and toward different directions within a galaxy, along with the possibility of mimicking photometric properties 
of stars \citep{Girardi2005, Vanhollebeke2009}. In this context, a consistent prediction of physical and 
photometric properties of stellar populations can put constraints on different evolutionary scenarios, and help in 
validating or ruling out viable theoretical models, when predictions are compared with observations.

Our Galaxy is a crucial laboratory where the validity of different models of stellar evolution and 
chemical enrichment in comparison with observations can be rigorously assessed. 
Observations especially in the solar neighbourhood and in the Milky Way (MW), but also 
in a limited number of nearby galaxies in the Local Universe provide us with large catalogues of stars, 
featuring accurate determinations of their chemical composition and stellar age, along with 
careful estimates of their physical and photometric properties \citep[e.g.][]{Stromgren1987, Edvardsson1993, 
Holmberg2000, Battinelli2003, Nordstrom2004, Melendez2008}.

However comparing these data with simulations in a fully consistent way is 
a non trivial and delicate task. 
State-of-the-art cosmological hydrodynamical simulations of both large cosmological volumes and 
individual galaxies have star particles whose mass typically ranges 
between $\sim$$10^8$ and $\sim$$10^3$ M$_{\odot}$ according to resolution
\citep{Dubois2014, Vogelsberger2014, Khandai2015, Rasia2015, Schaye2015, McCarthy2017, Remus2017, 
Pillepich2018, 
Oser2010, Aumer2013, Stinson2013, Marinacci2014, muppi2014, Dutton2015, Hopkins2017, Valentini2017}. 
In these simulations star particles are resolution elements and are treated as 
simple stellar populations (SSPs), i.e. ensembles of coeval stars that share the same initial metallicity. 

When comparing simulation predictions for the stellar metal content with observational data, 
the commonly pursued approach consists in analysing the total metallicity or different ion abundances that 
characterise the stellar mass of a system (e.g. a galaxy or a galaxy group), or of a portion of it. 
Such a method relies on the hypothesis that the mean metallicity of a star particle in simulations 
statistically reproduces the mean metal content of the SSP that the star particle samples.
A straightforward comparison with observational data is indeed fair when investigating the stellar metal content as 
a single quantity per galaxy, as for instance for the metallicity of stars as a function of the stellar mass 
of galaxies \citep{Schaye2015, Bahe2017, Dolag2017}, and for the stellar $\alpha$-enhancement 
as a function of galaxy stellar mass \citep{Segers2016}.

However, inconsistencies can arise when contrasting directly simulation output to observations of resolved stars. 
Surveys in the solar neighbourhood 
(\citealp[e.g. the Geneva-Copenhagen Survey,][]{Nordstrom2004}; \citealp[RAVE,][]{Steinmetz2006}; 
\citealp[SEGUE-1 and SEGUE-2,][]{Yanny2009}, \citealp{Rockosi2009}; 
\citealp[LEGUE,][]{Deng2012}; \citealp[the GAIA-ESO survey,][]{Gilmore2012}; 
\citealp[the AMBRE Project,][]{deLaverny2013}; \citealp[APOGEE,][]{Majewski2016}; 
\citealp[the SkyMapper Southern Survey,][]{Wolf2018})
and observations in Local Universe galaxies 
resolve indeed individual stars and provide us with accurate determinations of single star properties. 

Going from star particles in simulations to observed stars is therefore crucial to accurately compare 
simulation predictions with data. The key questions that we want to address with our study are the following: is it 
possible to bridge the gap between the typical mass of stellar particles in simulations and that of observed 
single stars? In other words, would that be achievable to develop a robust method that allows us to compare 
simulation outcome with observations of explicitly resolved stars?

In this paper we introduce a novel tool that allows to consistently generate a population of stars associated 
to a single star particle, which is characterised by a metallicity, age and initial mass function (IMF).  
This tool takes properties of star particles from simulations as input, 
along with their position and an IMF; a theoretical catalogue of mock stars 
whose characteristics are drawn from the input features is then generated, by using 
the TRILEGAL code \citep{Girardi2005}, that is able to reproduce the photometric properties of mock stars. 
The ultimate goal of this tool is to generate a database of synthetic stars from the properties 
of parent star particles in simulations: such a database can be deemed as the observable stellar content 
corresponding to observational data. 
It translates the populations of star particles of a simulation into stellar populations, providing masses, ages, 
metallicities and magnitudes of individual stars, thereby enabling a direct comparison with observations. 
The method is flexible, can be easily tailored to fit outputs of different codes used for cosmological simulations, and 
of paramount importance with ongoing survey data releases (e.g. GAIA and surveys of resolved stellar populations).

The outline of this paper is as follows. We give an overview of the simulations and describe the 
sub-resolution model that we use in Section \ref{sec2}. In Section \ref{Method} we introduce the methodology 
adopted, while Section \ref{Data} provides the main features of the observational sample that we choose to 
compare our results with. Our results are presented in Section \ref{Results}. We discuss our main findings in 
Section \ref{Discussion}, we summarise our key results and draw conclusions in Section \ref{sec:conclusions}.

\section{Cosmological simulations of a disc galaxy}
\label{sec2}
 
\subsection{Simulations}
\label{TheSim}

To introduce our tool we first perform simulations that provide an accurate model of a disc galaxy. 
Since careful observations are mostly available for our Galaxy, we conceived simulations of a galaxy that 
turns out to have similar properties to the MW ones.

We start our work by carrying out three cosmological hydrodynamical simulations with zoomed-in 
initial conditions (ICs) of an isolated 
dark matter (DM) halo of mass $M_{\rm halo, \, DM} \simeq 2 \cdot 10^{12}$~M$_{\odot}$ at redshift $z=0$. 
These ICs have been first introduced by \citet{Springel2008} at different resolutions and named $AqC$. 
The halo that we simulate is expected to host a late-type galaxy at redshift $z=0$, also because of 
its quite low-redshift merging history. 
Despite of the similarity between the simulated galaxies and our Galaxy 
from the morphological point of view, our results should not be considered as a model of the MW. 
In fact, no specific attempts to reproduce the accretion history of its dynamical environment have been made. 
The zoomed-in region that we simulate has been selected within a cosmological volume of 
$100 \, (h^{-1}$ Mpc$)^{3}$ of the DM only parent simulation. As for the cosmological model, 
we adopt a $\Lambda$CDM cosmology, with $\Omega_{\rm m}=0.25$, $\Omega_{\rm \Lambda}=0.75$, 
$\Omega_{\rm baryon}=0.04$, $\sigma_8 = 0.9$, $n_s=1$, 
and $H_{\rm 0}=100 \,h$ km s$^{-1}$ Mpc$^{-1}=73$ km s$^{-1}$ Mpc$^{-1}$.

The simulations have been carried out with the TreePM+SPH (smoothed particle hydrodynamics) GADGET3 code, 
a non-public evolution of the GADGET2 code \citep{Springel2005}. 
This version of the code uses the improved formulation of 
SPH presented in \citet{beck2015}. This SPH implementation includes 
a higher order kernel function, an artificial conduction term and a correction for the artificial viscosity (AV), thus 
ensuring a more accurate fluid sampling, an improved description of hydrodynamical instabilities, and a removal 
of AV away from shock regions. 
To describe a multiphase ISM and a variety of physical processes we used the sub-resolution model 
MUPPI \citep{muppi2010,muppi2014}, as detailed in Section \ref{sec:muppi}. 
The new SPH formulation has been introduced in cosmological simulations adopting MUPPI 
according to \citet{Valentini2017}. 

We adopt a Plummer-equivalent softening length for the computation of the gravitational interaction of 
$\varepsilon_{\rm Pl} = 325 \, h^{-1}$~pc. We assume the softening scale of the gravitational force to have 
a constant value in physical units up to $z=6$, and a constant value in comoving units at earlier epochs.
Mass resolutions for DM and gas particles are the following: DM particles have a mass of 
$1.6 \cdot 10^6 \, h^{-1}$~M$_{\odot}$, while the initial mass of gas particles is $3.0 \cdot 10^5 \, h^{-1}$~M$_{\odot}$. 
We note that the mass of gas particles is not constant throughout the simulation: the initial mass can indeed 
decrease due to star formation (i.e. spawning of star particles), and it can increase because of gas return 
by neighbour star particles. 

In this work we consider results of three simulations that we carried out: AqC5--fid (our reference simulation), 
AqC5--cone, and AqC5--3sIMF (see Table \ref{TableParamm}). 
The major differences between the simulations are the following: AqC5--fid and AqC5--cone share the same IMF 
(see below) but have a different galactic outflow model (see Section \ref{sec:muppi}); AqC5--fid and AqC5--3sIMF 
have the same galactic outflow model but are characterised by a different IMF. The main features of these simulations 
are outlined in Section \ref{featuresOfTheGalaxy}. Here, we anticipate that simulations AqC5--fid and AqC5--cone 
have been introduced in \citet{Valentini2017}, while a companion paper (Valentini et al., in preparation) will 
thoroughly address the role of IMF in simulations of disc galaxies.

Both the simulations AqC5--fid and AqC5--cone adopt a \citet{Kroupa2001} IMF that is characterised by two slopes 
(K2s, hereafter; this IMF is a variation of the \citealt{kroupa93} IMF). We define the IMF $\phi (m)$ as: 
\begin{equation}
\phi (m) = \beta m^{-(1+x)} =  \beta m^{- \alpha}
\label{IMF}
\end{equation}
in the mass range $[0.08, 100]$ M$_{\odot}$. The slope $\alpha$ of the power law has the following values 
in different mass intervals: 
\begin{equation}
\begin{array}{l}
\alpha = 1.3    $\,\,\,\,\,\,\,\,\,\,$ {\text {for}} $\,\,\,\,\,\,$ 0.08 {\text { M}}_{\odot} \le m \le 0.5 {\text { M}}_{\odot}, \\
\alpha = 2.3    $\,\,\,\,\,\,\,\,\,\,$ {\text {for}} $\,\,\,\,\,\,$ 0.5 {\text { M}}_{\odot} < m \le 100 {\text { M}}_{\odot}.  \\
\label{IMFonly2slopes}
\end{array}
\end{equation}
Note the use of the IMF exponent $\alpha = 2.3$ for massive stars, not corrected for unresolved stellar binaries 
(see below).
For each mass span a normalization constant $\beta$ is computed by imposing that 
$\int m \, \phi (m) \, dm = 1$ over the global mass range and continuity at the edges of subsequent mass intervals.

The third simulation, AqC5--3sIMF, is a simulation that is similar to the reference AqC5--fid, but for the adopted IMF. 
In AqC5--3sIMF we adopt the \citet{kroupa93} IMF, as suggested by \citet{Grisoni2017}. This IMF is characterised 
by three slopes (K3s, hereafter) and is defined in the mass range $[0.1, 100]$ M$_{\odot}$. 
For the latter simulation the slope $\alpha$ of the power law in equation (\ref{IMF}) has the following values 
according to the mass interval:
\begin{equation}
\begin{array}{l}
\alpha = 1.3    $\,\,\,\,\,\,\,\,\,\,$ {\text {for}} $\,\,\,\,\,\,$ 0.1 {\text { M}}_{\odot} \le m \le 0.5 {\text { M}}_{\odot}, \\
\alpha = 2.2    $\,\,\,\,\,\,\,\,\,\,$ {\text {for}} $\,\,\,\,\,\,$ 0.5 {\text { M}}_{\odot} < m \le 1.0 {\text { M}}_{\odot},  \\
\alpha = 2.7    $\,\,\,\,\,\,\,\,\,\,$ {\text {for}} $\,\,\,\,\,\,$ 1.0 {\text { M}}_{\odot} < m \le 100 {\text { M}}_{\odot}.  \\
\label{IMFslopes}
\end{array}
\end{equation}
In this case, the slope $\alpha = 2.7$ for massive stars includes the correction for unresolved 
stellar binary systems.

\begin{table*}
\centering
\begin{minipage}{175mm}
\caption{Relevant parameters of the sub-resolution model.
Column 1: simulation name.
Column 2: adopted IMF. K2s is Kroupa IMF with two slopes, while K3s is Kroupa IMF with three slopes (see the text).
Column 3: temperature of the cold phase.
Column 4: pressure at which the molecular fraction is $f_{\rm mol}=0.5$.
Column 5: number density threshold for multi-phase particles.
Column 6: gas particle's probability of becoming a wind particle.
Columns 7: half-opening angle of the cone for thermal feedback, in degrees. As for AqC5-cone, this is 
	also the half-opening angle of the cone for kinetic feedback.
Columns 8 and 9: thermal and kinetic SN feedback energy efficiency, respectively.
Column 10: fraction of SN energy directly injected into the hot phase of the ISM.
Column 11: evaporation fraction. 
Column 12: star formation efficiency, as a fraction of the molecular gas.
Column 13: number of stellar generations, i.e. number of star particles generated by each gas particle.
Column 14: average stellar masses of stars formed per each SN~II.}
\renewcommand\tabcolsep{2.9mm}
\begin{tabular}{@{}lccccccccccccc@{}}
\hline
Name & IMF & $T_{\rm c}$ & $P_{\rm 0}$ & $n_{\rm thresh}$ 
& $P_{\rm kin}$ &  $\theta$  & $f_{\rm fb, therm}$ & $f_{\rm fb, kin}$ 
& $f_{\rm fb, local}$ & $f_{\rm ev}$ & $f_{\star}$ & $N_{\ast}$ & $M_{\ast, \rm SN}$ \\ 
& & $(K)$ & (k$_{\rm B}$K cm$^{-3}$) & (cm$^{-3}$) 
&  & ($^{\circ}$) & &  
&  &  & & & (M$_{\odot}$)  \\ 
\hline
\hline
AqC5--fid & K2s &  300 & 2 $\cdot 10^4$ &  0.01 & 0.03 & 30 & 0.2 & 0.12 
& 0.02 & 0.1 & 0.02 & 4 & 120\\  
\hline
AqC5--cone & K2s &  300 & 2 $\cdot 10^4$ &  0.01 & 0.05 & 30 & 0.2 & 0.7 
& 0.02 & 0.1 & 0.02 & 4 & 120\\  
\hline
AqC5--3sIMF & K3s &  300 & 2 $\cdot 10^4$ &  0.01 & 0.03 & 30 & 0.2 & 0.26 
& 0.02 & 0.1 & 0.02 & 4 & 120\\  
\hline
\hline

\end{tabular}
\label{TableParamm}
\end{minipage}
\end{table*}

We account for different timescales of evolving stars with different masses by adopting the mass-dependent 
lifetimes by \citet{PadovaniMatteucci1993}. 
We consider $8$~M$_{\odot}$ as the minimum mass giving rise to stellar BHs, and that stars that are more massive 
than $40$ M$_{\odot}$ implode in BHs directly, thus not contributing to further chemical enrichment.

We assume that a fraction of stars relative to the entire mass range is located in binary systems suitable 
for being progenitors of SNe~Ia. Such a fraction is $0.03$ \citep[as suggested by][]{Grisoni2017} in 
the simulation AqC5--3sIMF; this parameter was set to 0.1 in simulations AqC5--fid and AqC5--cone.
The binary system producing a SN Ia must have a total mass varying from $3$ to $16$~M$_{\odot}$. 
The first SNe~Ia explode therefore in binary systems where both stars have initial mass of $8$~M$_{\odot}$, 
according to the single-degenerate scenario. This sets the timescale required for SN~Ia explosions and 
consequent chemical enrichment, that ranges between $\sim 35$ Myr and several Gyr.

The production of different heavy metals by aging and exploding stars is followed by assuming 
accurate sets of stellar yields. We adopt the stellar yields provided by \citet{Thielemann2003} for SNe~Ia and 
the mass- and metallicity-dependent yields by \citet{Karakas2010} for intermediate and low mass stars 
that undergo the AGB phase.
As for SNe~II, we use the mass- and metallicity-dependent yields by \citet{WoosleyW1995}, 
combined with those provided by \citet{Romano2010}.
This set of yields has been tested by state-of-the-art chemical evolution models and well reproduces observations 
of different ion abundances in the MW \citep[][and private communications]{Romano2010}.  
Relevant parameters adopted in the simulation are listed in Table \ref{TableParamm}.

We note that gas and stellar metal content and distribution are genuine predictions of our simulations, 
since our sub-resolution model has not been fine-tuned to reproduce observations of metal abundances 
in the ISM and CGM \citep[see e.g. the metal loading of wind particles in][]{Vogelsberger2013, Pillepich2018}.

\subsection{Modelling star formation and chemical enrichment}
\label{sec:muppi}

Cosmological hydrodynamical simulations for the formation of cosmic structures account for the joint evolution 
of both DM and baryons in a self-consistent way. Since galaxy formation involves a variety of phenomena 
developing over a large dynamic range, cosmological simulations resort to sub-resolution models to include 
processes occurring at scales below the resolution limit of the simulation.

We adopt a sub-resolution model called MUPPI (MUlti Phase Particle Integrator). In this section
we outline its most relevant features, while we refer the reader to \citet{muppi2010, muppi2014} and 
\citet{Valentini2017} for a more comprehensive description and for further details.
Our sub-resolution model describes a multi-phase interstellar medium (ISM), featuring star formation and 
stellar feedback, metal cooling and chemical evolution, and accounting for the presence of an ionizing 
background. The building block of the model is the multi-phase particle: it consists of a hot and a cold 
gas phases in pressure equilibrium, and a possible further stellar component. 
A gas particle enters a multi-phase stage whenever its density increases above a density threshold 
and its temperature drops below a temperature threshold ($T_{\rm thresh}=10^5$ K). 
The number density threshold is $n_{\rm thres}=0.01$ cm$^{-3}$, that corresponds to 
a number of hydrogen particles $n_{\rm H} \sim 0.0045$ cm$^{-3}$ 
(adopting $0.76$ as the fraction of neutral hydrogen), and to 
a density $\rho_{\rm thres} \simeq 1.5 \cdot 10^{-4}$ M$_{\odot}$ pc$^{-3}$ 
(assuming $\mu \sim 0.6$ as mean molecular weight). 
When a gas particle becomes a multi-phase particle, 
it is considered to be made of hot gas only.  
This hot component then cools down according to its density and metallicity,
thus generating the cold component of the multi-phase particle, whose temperature is fixed to 
$T_{\rm c}=300$ K.

Mass and energy flow among different components according to a set of ordinary
differential equations. 
Hot gas condenses into a cold phase due to radiative cooling, while a tiny part $f_{\rm ev}$ of the cold gas 
in turn evaporates because of destruction of molecular clouds (we list in Table \ref{TableParamm} the values 
for the model's parameters adopted in our simulations). 
A fraction $f_{\rm mol}$ of the cold gas mass $M_{\rm c}$ is expected to be in the molecular phase: molecular 
gas is then converted into stars according to a given efficiency ($f_{\ast}$). The SFR (star formation rate) 
associated to a multi-phase particle is therefore:
\begin{equation}
\centering
\dot{M}_{\rm sf} \:=\: f_{\ast} \: \frac{f_{\rm mol} \, M_{\rm c}}{t_{\rm dyn}} \,.
\label{eq:sfr}
\end{equation}
\noindent
Here, $t_{\rm dyn}$ is the dynamical time of the cold phase. The molecular fraction $f_{\rm mol}$ is
computed according to the phenomenological prescription by \citet{blitz2006}:
\begin{equation}
\centering
f_{\rm mol} \:=\: \frac{1}{1+P_0/P} \,\,, 
\label{eq:f_mol}
\end{equation}
\noindent
where $P$ is the hydrodynamic pressure of the gas particle and the parameter $P_0$, the pressure 
of the ISM at which $f_{\rm mol}=0.5$, is derived from observations (see Table \ref{TableParamm}). 

Star formation is modelled according to the stochastic scheme introduced by 
\citet{SpringelHernquist2003}. A multiphase gas particle with initial mass $M_{\rm gas, init}$ 
can generate a star particle of mass $M_{\ast, \rm init}$ with a probability: 
\begin{equation}
\centering
p \:=\: \frac{M_{\rm gas, init}}{M_{\ast, \rm init}} \Biggl[ 1 - {\text {exp} } \Biggl( - \frac{\Delta M_{\ast}}{M_{\rm gas, init}}  \Biggr) \Biggr] \,\,, 
\label{eq:SF}
\end{equation}
\noindent
where $\Delta M_{\ast}$ is the mass of the multiphase particle that has been
converted into stars in a time-step according to equation (\ref{eq:sfr}). 
Each star particle is spawned with mass $M_{\ast, \rm init} = M_{\rm gas, init}/N_{\ast}$, 
$N_{\ast}$ being the number of stellar generations. 
This number is a free parameter of the model: we choose $N_{\ast}=4$ in order to have an accurate 
representation of the star formation process, but no significant variations are observed 
for small deviations from this number \citep{tornatore2007}.

Radiative cooling is counterbalanced by the energy contributed by stellar feedback (see below)
and the hydrodynamical term accounting for
shocks and heating or cooling due to gravitational compression or expansion of gas.

We account for stellar feedback both in thermal and kinetic forms. 
As for thermal feedback \citep{muppi2010}, 
our model considers that the hot gas component of multiphase particles 
is heated both by the energy released directly by SN explosions related to the stellar component of 
the particle itself, and by energy contribution from surrounding SN explosions, i.e. neighbouring star-forming
particles contribute to the energy budget of each multiphase particle.
Therefore, a fraction $f_{\rm fb, local}$ of $E_{\rm SN} = 10^{51}$ erg is deposited directly 
in the hot gas component of multiphase particles, $E_{\rm SN}$ being the energy supplied by each SN.
Moreover, each star-forming particle provides neighbour particles with the following amount of thermal energy
in a given time-step: 
\begin{equation}
\centering
\Delta E_{\rm fb, therm}= f_{\rm fb, therm} \: E_{\rm SN} \frac{\Delta M_{\ast}}{M_{\ast, \rm SN}}\,.
\label{eq:thFB}
\end{equation}
Here, $f_{\rm fb, therm}$ is the thermal feedback efficiency, 
$M_{\ast, \rm SN}$ describes the stellar mass that is required on average to have a single SN~II, 
and $\Delta M_{\ast}$ is the mass of the multi-phase particle that has been converted into stars. 
Each star-forming particle provides this amount of thermal feedback energy to those neighbours 
that are located within a cone whose half-opening angle is $\theta$. The cone originates
on the star-forming particle itself and its axis is aligned according to (minus) the particle's 
density gradient. Energy contributions to eligible particles are weighted according to the SPH kernel, 
where the distance from the axis of the cone is considered instead of the distance between particle pairs.
If there are no particles in the cone, the total amount of thermal energy is given to the particle 
that lies nearest to the axis \citep{muppi2010,muppi2014}. 

The kinetic stellar feedback is responsible for triggering galactic outflows and it is modelled as follows 
\citep{Valentini2017}.
ISM is isotropically provided with kinetic feedback energy in simulations AqC5--fid and AqC5--3sIMF. 
Therefore, each star-forming particle supplies the energy
\begin{equation}
\centering
\Delta E_{\rm fb, kin}= f_{\rm fb, kin} \: E_{\rm SN} \frac{\Delta M_{\ast}}{M_{\ast, \rm SN}}\,
\label{eq:kinFB}
\end{equation}
to all wind particles (see below) within the smoothing length, with kernel-weighted contributions. 
In equation (\ref{eq:kinFB}), $f_{\rm fb, kin}$ describes the kinetic stellar feedback efficiency. 
Wind particles receiving energy use it to boost their velocity along their least resistance
path, since they are kicked against their own density gradient.
In simulation AqC5--cone the kinetic feedback energy is provided to wind particles located within a cone, 
whose half-opening angle is $\theta$, aligned towards the least resistance path of the energy donor particle. 
In this case particles are launched toward the direction opposite to the density gradient of the star-forming 
particle that provides them with feedback energy.

A gas particle exits the multiphase stage whenever its density drops
below $0.2 \rho_{\rm thresh}$ or after a maximum time interval (that
is set by the dynamical time of the cold gas). 
Should a gas particle be eligible to quit a multi-phase stage, it has a probability
$P_{\rm kin}$ of being kicked and to become a wind particle for a
time interval $t_{\rm wind}$. Both $P_{\rm kin}$ and $t_{\rm wind}$
are free parameters of the outflow model. This numerical procedure relies on the
physical idea that stellar winds are fostered by SN~II explosions,
once molecular clouds out of which stars formed have been
destroyed. 
Wind particles are decoupled from the surrounding medium
for the aformentioned lapse of time $t_{\rm wind}$. Despite being decoupled for this
time interval, wind particles receive kinetic energy from neighbouring star-forming gas
particles, as described above. The wind stage can be
concluded before $t_{\rm wind}$ whenever the particle density drops below a
chosen density threshold, $0.3 \rho_{\rm thresh}$, 
meaning that a wind particle has finally gone away from star-forming regions.

Besides the stellar feedback in energy, star formation and evolution also generate 
a chemical feedback, and galactic outflows foster metal spread and circulation throughout the galaxy.
Our model self-consistently accounts for the chemical evolution and enrichment processes, 
following \citet{tornatore2007}, where a thorough description can be found. 
Here we only highlight the most crucial features of the model.

\begin{figure}
\newcommand{\captionfonts}{\small}
\hspace{-2.1ex}
\includegraphics[trim=0.1cm 0.2cm 0.0cm 0.1cm, clip, width=0.5\textwidth]{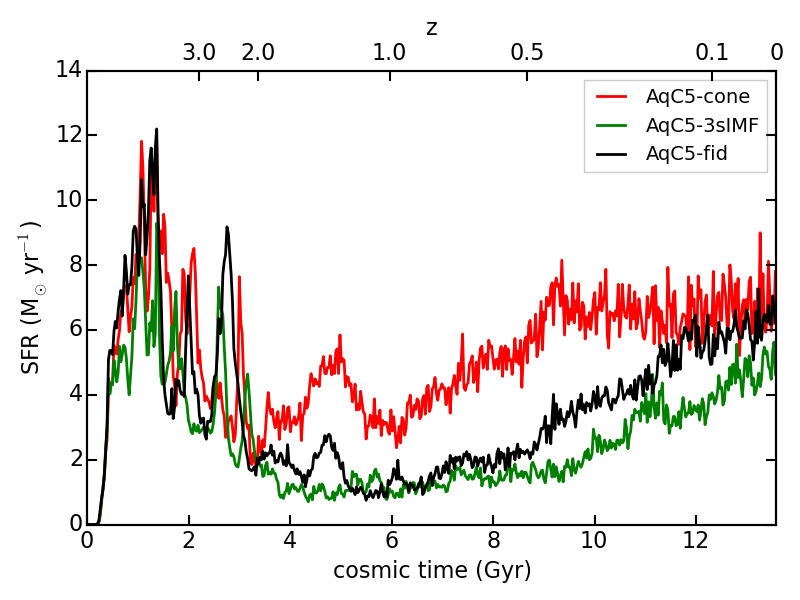} 
\caption{Star formation histories of the three simulated galaxies: our reference simulation in shown in black, 
red and green curves describe the evolution of the star formation rate (SFR) for AqC5--cone and AqC5--3sIMF, 
respectively.}
\label{sfr} 
\end{figure}

Each star particle initially shares the chemical composition of the gas particle from which it has been originated. 
Star particles are considered to be SSPs. By assuming an IMF 
and adopting accurate predictions for stellar lifetimes and stellar yields (see Section \ref{TheSim} for details), 
we carefully evaluate the number of stars aging and eventually exploding as SNe, as well as the amount of metals 
polluting the surrounding ISM. Different heavy elements produced and released by star particles are distributed 
to neighbouring gas particles, so that subsequently generated star particles are richer in heavy metals. 
The chemical evolution process is therefore responsible for the gradual reduction of the initial mass 
of stellar particles, too. 
We follow in details the chemical evolution of $15$ elements (H, He, C, N, O, Ne, Na, Mg, Al, Si, S, Ar, Ca, Fe and Ni) 
produced by different sources, namely AGB (asymptotic giant branch) stars, SNe~Ia and SNe~II.
Each atomic species independently contributes to the cooling rate, that is implemented 
according to \cite{wiersma2009}. When computing cooling rates, we also include the effect of 
a spatially uniform, redshift-dependent ionizing cosmic background \citep{HaardtMadau2001}.

\begin{figure*}
\newcommand{\captionfonts}{\small}
\begin{minipage}{\linewidth}
\centering
\includegraphics[trim=8.0cm 5.cm 10.0cm 3.5cm, clip, width=0.492\textwidth]{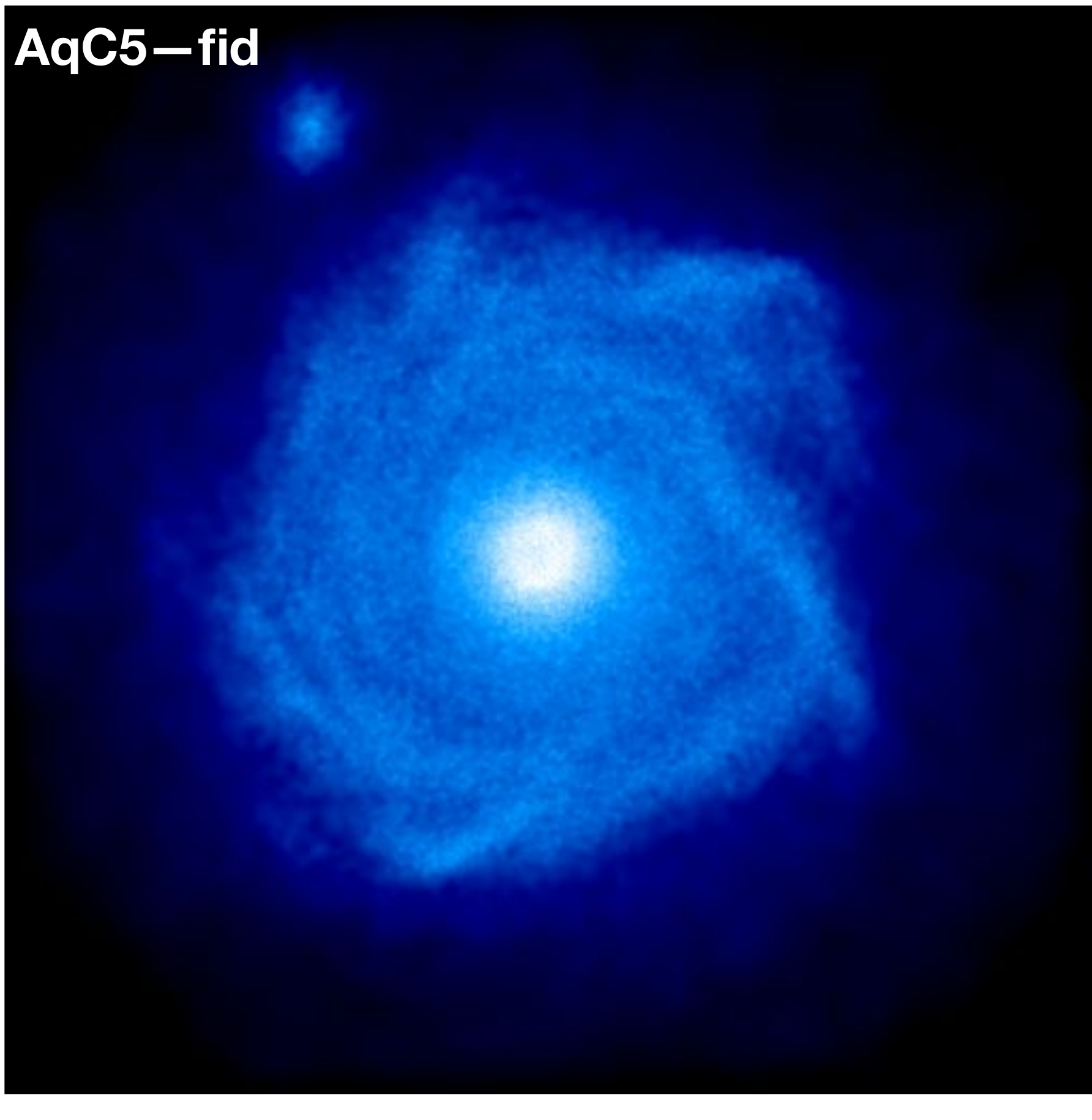} 
\includegraphics[trim=8.0cm 4.5cm 10.0cm 4.cm, clip, width=0.492\textwidth]{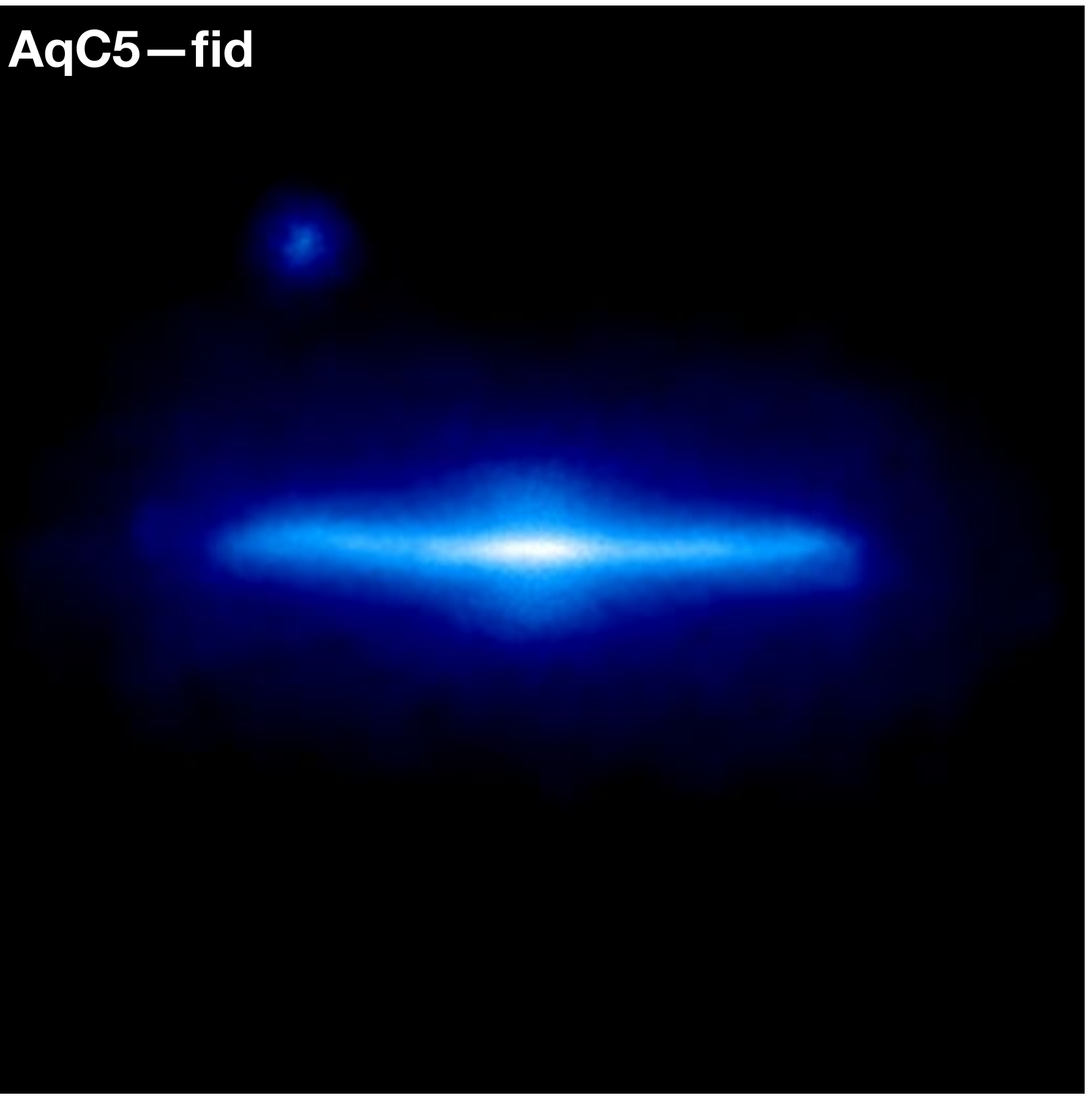} 
\includegraphics[trim=9.cm 4.0cm 10.0cm 3.5cm, clip, width=0.49\textwidth]{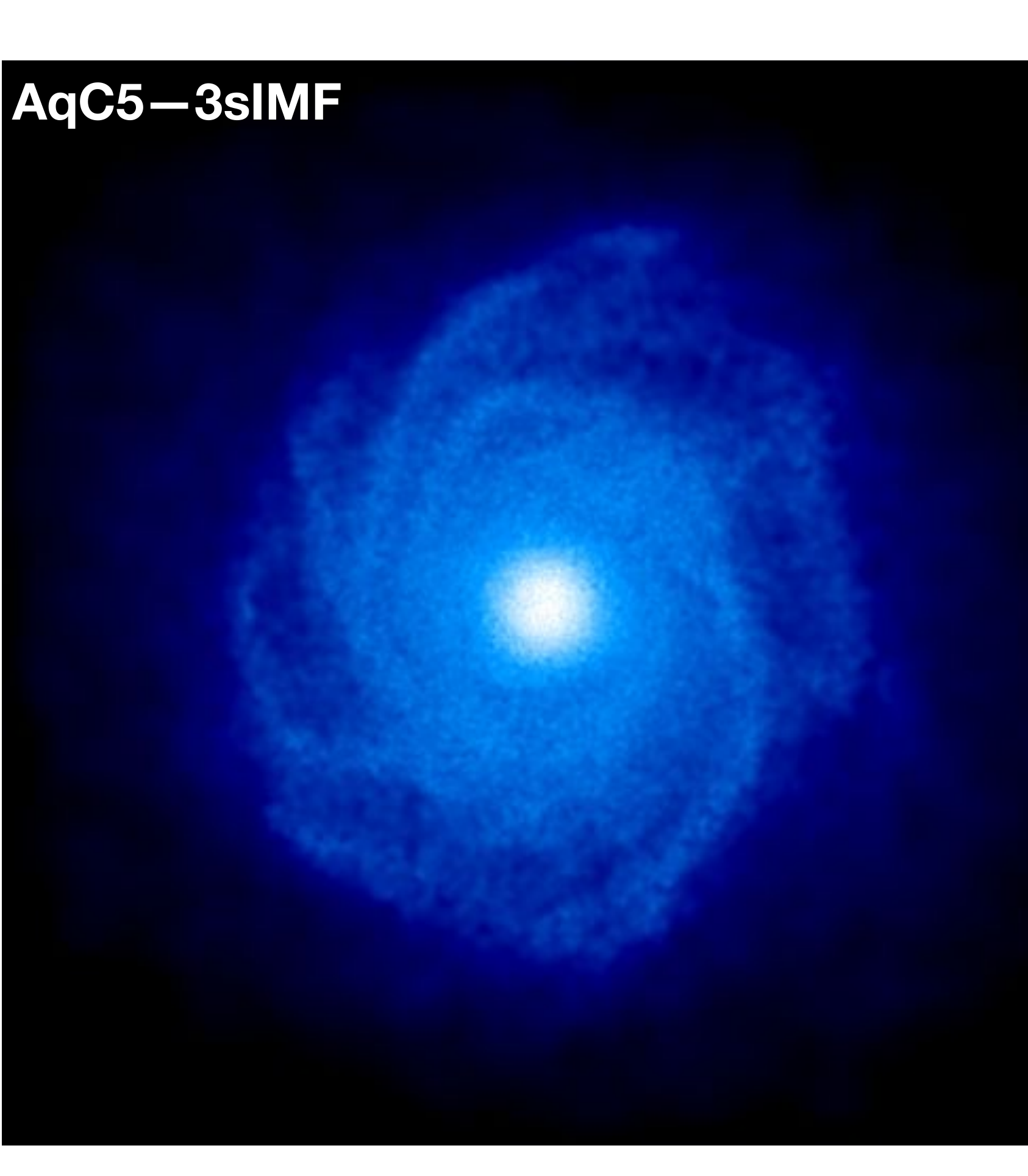} 
\includegraphics[trim=9.cm 4.0cm 10.0cm 3.5cm, clip, width=0.49\textwidth]{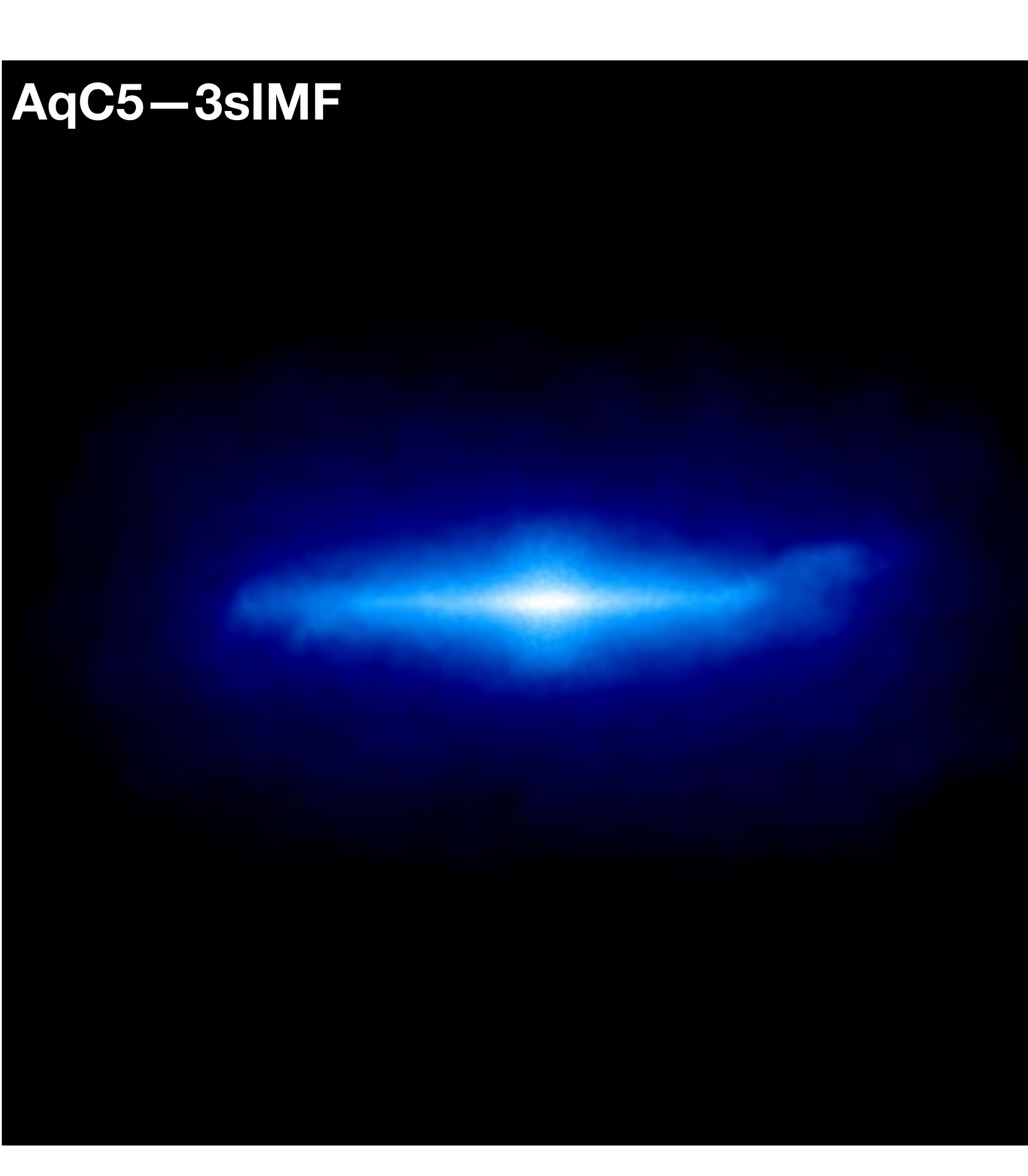} 
\end{minipage} 
\caption{Projected stellar density maps for the simulated galaxies AqC5--fid (top panels) and AqC5--3sIMF 
(bottom panels) at redshift $z = 0$. Left- and right-hand panels 
show face-on and edge-on densities, respectively. The size of the box is 40 kpc a side. 
Stellar density maps of the galaxy AqC5--cone, that we also consider for our analysis in Section \ref{otherResults}, 
are shown in \citet[][figure 6, third and fourth panels]{Valentini2017}.}
\label{StellarDensityMaps} 
\end{figure*}

\subsection{Main features of the simulated galaxies}
\label{featuresOfTheGalaxy}
We introduce the main physical properties of the galaxies resulting from our cosmological simulations. 
Galaxies AqC5--fid and AqC5--cone have been extensively analysed in 
\citet[][where they were labelled AqC5--FB2 and AqC5--newH, respectively]{Valentini2017}. 
We refer the interested reader to the aformentioned paper, where properties such as 
the gas mass accretion history within different radii, rotation curve and baryon conversion efficiency, 
density and metallicity profiles of these galaxies have been thoroughly investigated. 
Since this work is aimed at presenting a novel approch that can be used regardless of the peculiar features of a 
simulated galaxy, providing a detailed description of all the characteristics of our galaxies is beyond the scope of the 
present methodological paper. Here we only outline the properties of the AqC5--3sIMF galaxy that are useful to 
understand and interpret the results that we will discuss in Section \ref{otherResults}. A complete description 
of the properties of this galaxy will be given in a companion paper (Valentini et al., in preparation).

In Figure \ref{sfr} we show the star formation history of the three galaxies. The bulk of the stellar mass 
in the bulge of each galaxy builds up at high-redshift ($z \gtrsim 3$), as a consequence of a star formation burst. 
Star formation in the disc occurs at later epochs, in a more continuous way. AqC5--fid and AqC5--3sIMF 
are characterised by a lower SFR below $z \lesssim 2$ with respect to AqC5--cone, as a consequence of 
the reduced amount of gas that has been accreted within the galactic radius\footnote{We define here the galactic 
	radius as one tenth of the virial radius, i.e. $R_{\rm gal}= 0.1 R_{\rm vir}$. We choose 
  	the radius $R_{\rm gal}$ so as to identify and select the region of the 
  	computational domain that is dominated by the central galaxy.
  	Moreover, we consider virial quantities as those computed in a sphere that is centred 
	on the minimum of the gravitational potential of the halo and that encloses 
	an overdensity of 200 times the {\sl critical} density at present time.}.
The low-redshift ($z \lesssim 1.5$) SFR of AqC5--3sIMF is $\lesssim 5$ M$_{\odot}$ yr$^{-1}$, 
in agreement with the estimate provided by \citet{Snaith2014} for our Galaxy. 

Figure \ref{StellarDensityMaps} shows face-on (left-hand panels) and edge-on (right-hand panels) 
projected stellar density maps of star particles within the galactic radius of AqC5--fid (top panels) and 
AqC5--3sIMF (bottom panels). Stellar density maps of the galaxy AqC5--cone are shown in 
\citet[][figure 6, third and fourth panels]{Valentini2017}. 
The galaxy AqC5--3sIMF has a galactic radius $R_{\rm gal} = 24.0295$ kpc, at redshift $z=0$. 
It has a limited bulge component and a dominant disc, where spiral arms are evident. 
Although defining the extent of the gaseous and stellar disc is a difficult task, we note that the stellar surface 
density declines outwards down to $15$ M$_{\odot}$~pc$^{-2}$ at a distance of $\sim 11$~kpc from the galaxy 
centre, whereas the gas surface density decreases to that same value when the radius is $\sim 15$~kpc.
The galaxy is characterised by a rotation curve that is not centrally peaked and that is flat at large radii, the 
circular velocity at $r=8$~kpc being $v_c \simeq 220$~km~s$^{-1}$.

Here, we list some of the global properties of the AqC5--3sIMF galaxy. The virial radius of the galaxy at redshift $z=0$ is 
$R_{\rm vir} = 240.295$~kpc. DM, stellar and gas masses within $R_{\rm vir}$ are $1.532 \cdot 10^{12}$~M$_{\odot}$,
$2.957 \cdot 10^{10}$~M$_{\odot}$, and $1.709 \cdot 10^{11}$~M$_{\odot}$, respectively. 
DM, stellar and gas masses within $R_{\rm gal}$ are $2.539 \cdot 10^{11}$~M$_{\odot}$,
$2.679 \cdot 10^{10}$~M$_{\odot}$, and $2.342 \cdot 10^{10}$~M$_{\odot}$, respectively. 
A kinematic decomposition of the stellar mass within $R_{\rm gal}$ based on the circularity of stellar orbits 
\citep{Scannapieco2009} yields $1.108 \cdot 10^{10}$~M$_{\odot}$ as the stellar mass in the bulge, 
$1.571 \cdot 10^{10}$~M$_{\odot}$ as the stellar mass rotationally supported and located in the disc, 
and a bulge over total stellar mass ratio $B/T = 0.41$. Analogous properties of AqC5--fid and AqC5--cone 
can be found in \citet{Valentini2017}.

\section{Building the stellar content of star particles: method}
\label{Method}

In this section we accurately detail all the steps that make up the methodology developed in this work. 
We use this identifying convention here and in the following: star particles in the simulation 
are referred to as {\sl{parent particles}}, while mock stars that are generated by each parent particle are 
{\sl{child particles}} or, simply, {\sl{stars}}.

Our method aims at generating the observational properties of the mock stars in a volume of a simulated galaxy. 
To achieve that, we produce for each parent particle of given mass, age, and metallicity a stellar population 
of that age and metallicity, and with the same total initial mass. 
For this purpose, 
we first select parent particles that meet specific criteria within the simulation output (Section \ref{solarNeigh}). 
These parent particles define an age-metallicity distribution that is exploited to generate a catalogue of 
synthetic stars (Sections \ref{Trilegal} and \ref{Generation of mock stars}). 
These stars are then displaced to a suitable distance and among them we extract the final catalogue of 
mock stars (Section \ref{CMD}) that will be compared with observations in Section \ref{Results}. 

Throughout this paper, we refer to metallicity as overall metal content $[Z/H]$. 
Here, $[Z/H]=$ log$_{10}(Z/H) -$ log$_{10}(Z/H)_{\odot}$ is the logarithm of the ratio of the abundance by mass 
of all elements heavier than helium ($He$ or $Y$) over hydrogen ($H$ or $X$), compared to that of the Sun. 
As for the present-day Sun's metallicity, we adopt $Z_{\odot}=0.01524 (\pm 0.0015)$ \citep{Caffau2011} and 
$Y_{\odot}=0.2485 (\pm 0.0035)$ \citep{Basu2004}, so that 
$(Z/X)_{\odot}=0.0207 (\pm 0.0015)$ and 
log$_{10}(Z/H)_{\odot} = -1.684$ \citep{Bressan2012}.

\subsection{Parent particle selection}
\label{solarNeigh}
We start by considering the output of our cosmological hydrodynamical simulation AqC5--fid described in 
Section \ref{TheSim}. 
We analyse the distribution of star particles at redshift $z=0$, i.e. the distribution that produces the smoothed 
stellar density maps analysed in Figure \ref{StellarDensityMaps}. As a first step of our post-processing analysis 
(and as done for the stellar density maps shown in Figure \ref{StellarDensityMaps}), we rotate the galaxy 
reference system so as the $z$-axis is aligned with the angular momentum of star particles and 
multiphase gas particles located within $8$ kpc from the minimum of the gravitational potential. The axes of 
the galaxy frame are therefore defined by the eigenvectors of the angular momentum tensor. The origin of 
the reference system is taken to be the centre of the galaxy, that is determined as the centre of mass of star particles 
and multiphase gas particles within $8$ kpc from the location of the minimum of the gravitational potential. 

We then select the parent particles for our analysis. Since the purpose of the present study is to introduce 
a tool that allows the comparison between the properties of stars predicted by our simulations and observed stars, 
we identify a volume resembling the solar neighbourhood in our simulation snapshot. 
Star particles located within this volume (the so called {\sl{selected region}}) will be considered for the construction 
of our parent particle sample. 

The selected region in the simulated galaxy is a sphere centred on the galactic plane at a distance 
$r_{\rm Sun} = 8.33$ kpc from the galaxy centre. Such a distance can be deemed as the distance of the Sun 
from the Galactic Centre of our Galaxy, according to state-of-the-art determinations 
\citep{Gillessen2009, Bovy2009}. 
We locate the centre of the sphere on the plane where the $z$ coordinate is $z=0$.
We verified that the angular distance from the x-axis, $\vartheta_{\rm Sun}$, does not affect final results 
(see Section \ref{Dipendenze}).

To estimate the size of the sphere, we determine its radius as follows. 
The actual resolution of our simulation amounts to $2.8$ times the formal resolution described by the 
gravitational softening \citep{Springel2001}, and hence slightly exceeds $1$~kpc 
($\varepsilon_{\rm Pl} = 325 \, h^{-1}$ pc, see Section \ref{TheSim}). 
Since we cannot rely on the sampling of the statistical properties of particles in smaller volumes, that represent our 
smaller possible solar neighbourhood (star particles keep their own properties, that have been set where and when 
they were originated). 
To illustrate how this size compares with the observational requirements, we compute the distance modulus 
$d_{[\rm pc]}$ of the brightest stars that can be seen within an apparent-magnitude limited survey. 
The associated distance defines a tentative size for the solar neighbourhood. 
If we consider the Geneva-Copenhagen Survey (see Section \ref{Data}), with an apparent-magnitude limit in 
the Johnson $V$ passband $m_V \sim 8.3$ and an absolute magnitude $M_V \sim 1$ 
\citep[from figure $12$ of ][]{Casagrande2011}, we end up with an estimate of a characteristic radius of the survey 
that is $d = 288$ pc\footnote{We 
		notice that if we adopted $m_V \sim 7.6$, that is the apparent magnitude of the Geneva-Copenhagen 
		Survey where incompleteness sets (see Section \ref{Data}), we would end up with a distance modulus 
		$d = 209$ pc. However, the procedure is not sensitive to the exact values assumed for $m_V$ and $M_V$.}. 

Selecting a sphere of radius $1$~kpc, we maximise the sampling of the statistical quantities that determine our results, while using a proper volume accordingly with our simulations' properties.

The parent particles selected for our analysis are all the star particles residing within the aformentioned volume 
of the simulated galaxy that can be compared to the solar neighbourhood. 
The typical number of star particles located in this volume is $\sim 1000$. For instance, in the reference selected sphere 
of AqC5--fid centred at $r_{\rm Sun} = 8.33$ kpc and $\vartheta_{\rm Sun}=60 ^{\circ}$ we have $891$ parent particles. 
When considering the different volumes analysed in Section \ref{Dipendenze} for AqC5--fid we find a mean value 
(and standard deviation) of $1018 (\pm 186)$ parent particles. The number of star particles depends on the stellar 
mass of the simulated galaxy, and marginally varies with the position on the galactic plane.

Each stellar particle is treated as a parent particle characterised by the following properties: 
\begin{itemize}
\item the initial mass, that is the mass that the star particle had when it was generated;
\item the current mass, i.e. the mass of the parent particle when the analysis is performed (that is at redshift $z=0$, in our case)\footnote{We remind the reader that our star particles lose mass as they evolve as SSPs, 
	see Section \ref{sec:muppi}.};
\item the age;
\item the distance from the centre of the selected region;
\item the initial mass fraction of metals. 
\end{itemize}

We note that the fraction of mass in metals represents the metallicity (i.e. $Z$ linear and absolute, meaning that 
it is not scaled according to the solar metallicity $Z_{\odot}$) of the parent particles, that is the metallicity that 
star particles had when they were originated. 
Along with this quantity, we also compute $[Z/H]$ for each parent particle: this is the way to express the metallicity 
that can be most easily related to data\footnote{We remind the reader that we know 
the mass in hydrogen, helium and other ions for each parent particle in our simulation.}.

\subsection{Generation of synthetic stars: the TRILEGAL Code}
\label{Trilegal}

The second building block that makes up our tool consists in using the TRILEGAL code  
to produce a catalogue of mock stars. 

The TRILEGAL code \citep[TRIdimensional modeL of thE GALaxy,][]{Girardi2005} is a population-synthesis code.
It produces mock stellar catalogues for a parent object (i.e. for each star particle, in our case) with given 
characteristics and simulates the corresponding photometric properties of the generated stars. 
The key goals of the TRILEGAL code are indeed to predict the expected star counts in different photometric systems 
and to mimic the photometric properties of stars located towards a given direction.
Interesting examples of how this tool can be used to investigate the characteristics of stars in the bulge of 
MW as well as in external galaxies can be found in \citet{Vanhollebeke2009} and \citet{Girardi2007}, respectively. 
Here we detail the most relevant features of the TRILEGAL code for the current study. For further technical information, 
we refer the reader to the aformentioned papers.

The TRILEGAL code assembles collections of stars by assuming an IMF and defines their photometric properties 
by using a library of stellar evolutionary tracks. First, it performs a Monte Carlo simulation to produce stars 
according to input features, that are the age-metallicity distribution (AMD) and the star formation rate (SFR). 
The IMF, SFR and AMD determine mass, age and metallicity of each single synthetic star. 
The code then carries out an interpolation within grids of theoretical evolutionary models, i.e. within isochrone sets, 
so as to generate the absolute photometry and provide the absolute bolometric magnitude of mock stars. 
This latter value is hence converted to absolute magnitudes in different passbands by taking advantage of 
tables of bolometric corrections, that have been constructed from a library of syntethic spectra. The 
photometric system according to which magnitudes in different passbands are computed can be defined as 
an input requirement.

A further input specification in the TRILEGAL code deals with the geometry of the system where mock stars 
are generated, i.e. with its stellar density as a function of the position. This information is needed to retrieve 
the number counts of stars towards an element of given coordinates subtending a certain solid angle, and 
in a given bin of magnitude \citep[see][for more details]{Girardi2005}. The tool allows to choose either a defined 
component of our Galaxy (for instance the bulge, the thin or thick disc, the halo) or additional objects at known 
distance. We adopt the latter choice and we explain why in Section \ref{Generation of mock stars}. 

The TRILEGAL code generates a catalogue of stars as output. For each star, the following properties are provided: 
age, metallicity, initial mass, luminosity along with effective temperature, surface gravity, 
absolute bolometric magnitude, and absolute magnitudes in different passbands.
Although the TRILEGAL code lets optionally assume a known foreground interstellar absorption, in this study 
we rely neither on a detailed modelling of distribution of dust nor on absorption curves. 
We rather assume a simple distance-dependent dust extinction law and a resulting color excess, as 
explained in Section \ref{CMD}.

We describe in Sections \ref{Generation of mock stars} and \ref{CMD} how we use this tool by providing it with 
properties of star particles from simulations as input, and how mock photometric properties of generated stars are 
used to derive a colour-magnitude diagram (CMD).

\subsection{Producing the star catalogue}
\label{Generation of mock stars}

Once the parent particles have been selected (see Section \ref{solarNeigh}), their properties are provided to 
the TRILEGAL code as input values, together with the adopted IMF, in the following way.

\begin{figure}
\newcommand{\captionfonts}{\small}
\begin{minipage}{\linewidth}
\centering
\hspace{-2.4ex}
\includegraphics[trim=0.2cm 0.0cm 0.0cm 0.0cm, clip, width=1.03\textwidth]{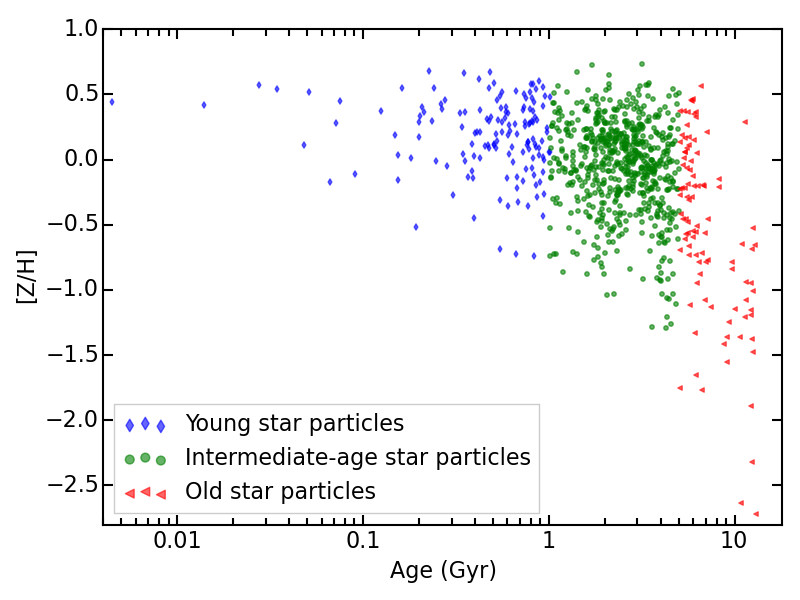} 
\end{minipage} 
\caption{Age-metallicity distribution for star particles divided into different age intervals, at redshift $z = 0$. 			Metallicity is 
	analysed in terms of $[Z/H]$. Star particles are colour-coded according to their age (see text): 
	blue diamonds show star particles younger than $1$ Gyr, 
	green circles depict star particles with age between $1$ and $5$ Gyr, 
	while star particles older than $5$ Gyr are identified by red triangles.}
\label{Age-metallicityRelation} 
\end{figure}

\begin{itemize}
\item We exploit the age of parent particles and their metallicity to provide the TRILEGAL code 
	with the AMD of star particles (see Figure \ref{Age-metallicityRelation} 
	and discussion below).
\item We require the TRILEGAL code to use the same IMF that we adopted in our cosmological simulation,
	that is a Kroupa IMF \citep[][either with two or three slopes according to the simulated galaxy that we 
	consider for our analysis, see Section \ref{TheSim}]{kroupa93}.
\item We further use the age and mass of parent particles to supply the TRILEGAL code with an SFR. 
\end{itemize}

The TRILEGAL code is hence provided with a table describing the number of parent particles, with given 
(initial and current) mass, per age and metallicity bin. We adopt logarithmic age bins spanning the range 
$0-13.56$ Gyr, in order to have a finer sampling of stellar ages at low redshift $z$ and a corser one at higher $z$. 
Metallicity bins are $0.05$ dex wide each. 
For instance, we can investigate age and metallicity properties of the parent particles within the volume centred 
at $r_{\rm Sun} = 8.33$ kpc and $\vartheta_{\rm Sun}=60 ^{\circ}$ for the reference AqC5--fid simulation. 
Figure \ref{Age-metallicityRelation} shows the AMD for all the star particles in this selected sphere. 
Parent particles are colour-coded according to their age: star particles younger than $1$ Gyr are shown in blue, 
parent particles having age between $1$ and $5$ Gyr are identified by green filled circles, whereas 
red triangles pinpoint stars older than $5$ Gyr. This age classification reflects the one adopted by 
\citet{Casagrande2011} and further discussed in Section \ref{otherResults}. 
We show the metal content of the star particles in terms of overall metal content $[Z/H]$.
Old parent particles have a spread in metallicity that is remarkably wider ($\sim 3$ dex) than younger star 
particles ($\lesssim 1.5$ dex), and the metallicity distribution progressively narrows as we approach the youngest ones.

As for the SFR, we assume an impulsive SFR for all the parent particles in each age and 
metallicity bin, so that we can deal with a SSP of that age and metallicity, and of given initial mass. 
The TRILEGAL code considers our parent particles as 
objects at known distance (see Section \ref{Trilegal}): each particle is located at a given distance from the centre 
of the sphere, that is the volume within which we perform our analysis.
We know indeed the position of each parent particle from the numerical simulation.

Each age and metallicity bin is then populated with stars. Since age and metallicity determinations in 
observations are affected by non negligible uncertainties, we generate stars for stellar populations whose age 
and metallicity are the ones of the bin. Different options are possible, such as the generation of stars for each 
star particle, i.e. for each stellar population with a peculiar age and metallicity. We prefer the first choice, that 
also has the advantage of speeding up the procedure and will be convenient in view of higher resolution 
simulations and larger-area coverage surveys.
While generating stars according to the input IMF, we choose to produce
as many stars as needed to reach a stellar mass of $M_{\ast \, \rm{Age, Z}} \simeq 10^5$ M$_{\odot}$ per 
age and metallicity bin. The TRILEGAL code requires the total stellar mass to be generated in each age 
and metallicity bin. We adopted this value after carrying out extensive tests. Our choice has been mainly driven 
by the fact that the stellar mass in each age and metallicity bin is at least as large as $\sim 10^5$ M$_{\odot}$, 
in case only a few star particles occupy an age and metallicity bin. 
$M_{\ast \, \rm{Age, Z}}$ is defined as $M_{\ast \, \rm{Age}} / N_{\rm Z \, bins}$, that is the stellar mass 
per age bin over the number of metallicity bins in the considered age bin. 
$M_{\ast \, \rm{Age, Z}}$ can be considered as the current default IMF normalization in each age and metallicity 
bin (see Section \ref{CMD}). 
We will only consider stars with mass larger than $0.6$ M$_{\odot}$ for the current analysis: the reason stems from 
the mass range of stars that are typically observed in the solar neighbourhood \citep[in particular, see 
Section \ref{Data} for the Geneva-Copenhagen Survey and][figure 16 $b$]{Casagrande2011}. 
Therefore, generated stars with mass lower than our adopted threshold contribute to the total budget of stellar mass 
produced per age bin, but are not taken into account for further analysis.
In this way, we end up with a catalogue of mock stars with simulated photometric properties as output.

\subsection{Construction of the CMD}
\label{CMD}

The most powerful tool to analyse a star cluster is the colour-magnitude diagram (CMD). A CMD is the 
observational plane where stars locate according to their apparent magnitude and their effective temperature, 
that is related to and expressed as the difference between magnitudes in two different bands, i.e. a colour. 
The final goal of this analysis is the construction of a CMD, starting from the star catalogue that we have generated.

Mock stars are produced with the following properties: mass, metallicity or $[Z/H]$, surface gravity, luminosity, 
effective temperature, absolute bolometric magnitude, and absolute magnitudes in the $V$ band and in the 
$uvby$ bands of the Str{\"o}mgren photometric system. We choose this system since it has been shown to be 
particularly accurate and suited for determining stellar properties from colour indices 
\citep[see e.g.][and Section \ref{Data}]{Casagrande2011}. 
The metallicity of stars is the $[Z/H]$ of the metallicity bin they belong to.

The next step towards the construction of the CMD is to identify the stars generated by each parent particle.
This procedure allows to assign to each star the appropriate distance from the observer, that is the distance 
of its parent particle from the centre of the sphere, and thus the corresponding apparent magnitude. 
Stars in a given age and metallicity bin whose metallicity $[Z/H]$ differs by $\pm 0.05$ dex at most 
from the $[Z/H]$ of a parent particle in the same age interval are identified as generated from that parent particle. 
The choice of $0.05$ dex reflects the typical observational uncertainty on metallicity estimates (see Section \ref{Data}).
While doing this assignment, we discard low surface gravity stars, i.e. stars with 
log~$g < 3.0$, to limit the sample to MS stars, as in the observational data set (see Section \ref{Data}; in Section 
\ref{Results} we will show the impact of the threshold surface gravity).
Stars in each age bin are in this way attributed to their own parent particles. 

TRILEGAL provides for each set (Age, Z) identifying bins an ensemble of stars generated from a 
stellar population whose {\sl{initial}} mass is $M_{\ast \, \rm{Age, Z}}$ and 
whose {\sl{current}} number of stars is $N_{\ast \, \rm{Age, Z}}$, the latter quantity neglecting stars 
with mass smaller than $0.6$~M$_{\odot}$ and surface gravity log~$g < 3.0$. 
A number $N_{\ast \, \rm{extr}}$ of stars is then attributed to each parent particle proportionally to 
its initial mass $M_{\rm parent, init}$, by randomly extracting among the $N_{\ast \, \rm{Age, Z}}$ stars\footnote{We 
	verified that the mass distribution of the randomly extracted stars is in agreement with the IMF.}. 
Therefore: 
\begin{equation}
N_{\ast \, \rm{extr}} \propto \frac{M_{\rm parent, init}}{M_{\ast \, \rm{Age, Z}}} \cdot N_{\ast \, \rm{Age, Z}} \,.
\label{Nextr}
\end{equation} 
Such a procedure accomplishes the mass normalization, overcoming the initial generation of a fixed 
stellar mass per age and metallicity bin, regardless of the actual number of parent particles 
in each age and metallicity interval. 
Moreover, in this way more massive parent particles contribute to the final sample with a 
larger number of stars. 

Once we populated star particles with stars, these stars have to be put to a suitable distance in order 
to construct a proper CMD.
Since the TRILEGAL code provides mock stars with absolute magnitudes, the distance modulus has to be 
computed in order to retrieve apparent magnitudes and construct the CMD. 
In keeping with the notion that hydrodynamical quantities in simulations are characterised by 
smoothing distances, we also express the location of stars around each parent particle with a 
probability distribution function. We assume for this distribution a Plummer profile (see below). In this way, first 
stars are assigned the distance $d_{\rm 0, parent}$ of their own parent particle from the centre 
of the selected sphere. We then displace stars marginally around their position before estimating 
the apparent magnitude: we take a random number in the range 
($d_{\rm 0, parent} - \varepsilon_d$, $d_{\rm 0, parent} + \varepsilon_d$) 
for the actual value of $d_{{\rm 0,} \ast}$, i.e. the distance of each star from the centre of the selected sphere. 

The random extraction is weighted according to the differential mass distribution $dM_{\rm Pl}(r)/dr$ 
relative to a Plummer profile, since the mass represented by a star particle in the simulation is assumed to be 
distributed over a non-zero volume through a Plummer density profile\footnote{Extensive tests show that 
	adopting other possible profiles for the mass distribution, e.g. constant weights corresponding to a uniform 
	distribution of stars around the star particle position, does not affect significantly the apparent CMD.}.
The density corresponding to the Plummer potential \citep{Plummer1911, BandT} is:
\begin{equation}
\rho_{\rm Pl}(r) = \frac{3 m}{4 \pi \varepsilon_{\rm Pl}} \biggl(1 + \frac{r^2}{{\varepsilon}_{\rm Pl}^2} \biggr)^{-5/2} \,,
\label{Plummer}
\end{equation}
where $m$ is the mass of the system, i.e. a parent particle in our case, needed for the normalization, 
and $r$ is the distance from the centre.
We use the Plummer-equivalent softening length for the gravitational interaction of our simulation 
for the Plummer scale length $\varepsilon_{\rm Pl}$ (see Section \ref{TheSim}). 
We fix $\varepsilon_d = 3 \cdot \varepsilon_{\rm Pl} = 1.335$ kpc.
Equation (\ref{Plummer}) is therefore used to retrieve $dM_{\rm Pl}(r)/dr$ and compute weights, 
by considering the fraction of mass of a parent particle within a given shell.
This is a key step, because it allows the association between parent particles and individual stars. 

We then compute $m_V$, i.e. the apparent magnitude in the Johnson $V$ passband:
\begin{equation}
m_V = M_V - 5 + 5 \cdot {\text{log}}_{10} \, d_{{\rm 0,} \ast \, \text{[pc]}} \,\,,
\label{app_magn}
\end{equation}
and introduce a distance-dependent dust extinction, so that:
\begin{equation}
m_{V, {\rm{obs}}} = m_V + 1 \text{ magn} \cdot \frac{d_{{\rm 0,} \ast \, \text{[kpc]}}}{1 \text{kpc}} \,.
\label{dust}
\end{equation}
We do not attempt to provide a self-consistent model for the distribution of dust in the solar neighbourhood here: 
we just consider an attenuation of $1$ magnitude per kpc \citep[see e.g.][]{Vergely1998}, in order to mimic the 
presence of dust. We adopt this choice because a negligible number of stars ends up to have 
an apparent magnitude larger than the magnitude limit of the survey due to dust extinction. 
However, refinements to the procedure are possible, also exploiting quantities provided by the simulation itself: 
for instance, a dust-to-gas mass ratio can be assumed or estimated from gas metallicity, 
a dust cross section can be adopted, and the gas density can be used to 
compute the optical depth, and thus the attenuation.

Then, we compute the colour $b-y$ by making the difference between the $b$ and $y$ magnitudes\footnote{We 
		note that the colour $b-y$ is remarkably similar to the colour $B-V$ of the Johnson photometry.} 
in the two passbands of the Str{\"o}mgren photometry.
We also consider the effect of reddening $E(b-y)$ on our colour index:
\begin{equation}
(b-y)_{\rm obs} = (b-y) + E(b-y) = (b-y) + 0.32 \text{ magn} \cdot \frac{d_{{\rm 0,} \ast \, \text{[kpc]}}}{1 \text{kpc}} \,.
\label{reddening}
\end{equation}
In equation (\ref{reddening}), the coefficient $0.32$ comes from the adopted extinction law \citep{Cardelli1989}. 

To allow a fair comparison with the observational data set, we then remove stars whose apparent magnitude 
$m_{V, {\rm{obs}}}$ exceeds the apparent-magnitude limit of the CGS.
When making the final star catalogue that we use to compare our predictions to observed properties, 
we select $6670$ stars, by randomly extracting out of our star collection the same number of stars that 
makes up the $irmf$ sample (see Section \ref{Data}) of the CGS.
The CMD is then accomplished.

\section{Observational data set for our comparison}
\label{Data}

A precise determination of observed star properties is crucial to validate models of chemical enrichment and 
predictions of heavy-metal abundance. The MW, and in particular the solar neighbourhood, are the places 
where this task can be most easily and accurately achieved.

In Section \ref{otherResults} we compare our simulation predictions to observational data of stars in the 
solar neighbourhood from the Geneva-Copenhagen Survey 
\citep[GCS hereafter,][]{Nordstrom2004, Holmberg2007, Holmberg2009, Casagrande2011}.
Here we summarise the most important features of this survey for the current work, referring the interested 
reader to the aformentioned papers for further details.

The CGS is one among the most comprehensive investigations of nearby stars in the solar neighbourhood. 
This survey is an apparent-magnitude limited survey, as it observes stars brighter than $m_V \sim 8$.
The CGS provides an ideal accurate database: ages, metallicities, kinematics and galactic orbits are 
determined for a sample of low-mass main sequence (MS) stars.
Observed stars have masses spanning the range $0.6-2.3$ M$_{\odot}$ and are long-lived objects: 
the age probability distribution for all stars in the GCS peaks around $2$ Gyr.
Therefore, the atmospheres of these dwarf MS stars carry key information on the chemical composition 
of the ISM out of which they formed in a certain position of the disc of our Galaxy.
The peaked age distribution \citep[see][figure 13]{Casagrande2011} reflects the bias that the CGS has 
towards very young and very old stars. A number of reasons contribute to that, such as the 
apparent-magnitude limit, and the removal of very blue and giant stars. 
Distances of observed stars have been determined based on $Hipparcos$ parallaxes, 
while their ages have been estimated by using BASTI and Padova isochrones 
\citep[][and references therein]{Pietrinferni2006, Bertelli2008, Bertelli2009}. 

Among the $\sim 17000$ stars observed in the CGS, we focus our attention on a sub-sample of them for our analysis, 
that is the so-called $irmf$ sample. It is a sample of $6670$ stars, containing stars with the best photometry.
Additionally, they are mostly single MS stars (surface gravity log $g \ge 3.0$).

Effective temperature and metallicity were determined using homogeneous Str{\"o}mgren photometry, 
being the $uvby$ photometric system \citep{Stromgren1963} well suited to derive 
stellar atmospheric parameters through different colour indices.

\begin{figure*}
\newcommand{\captionfonts}{\small}
\begin{minipage}{\linewidth}
\centering
\includegraphics[trim=0.0cm 0.0cm 0.0cm 0.0cm, clip, width=1.\textwidth]{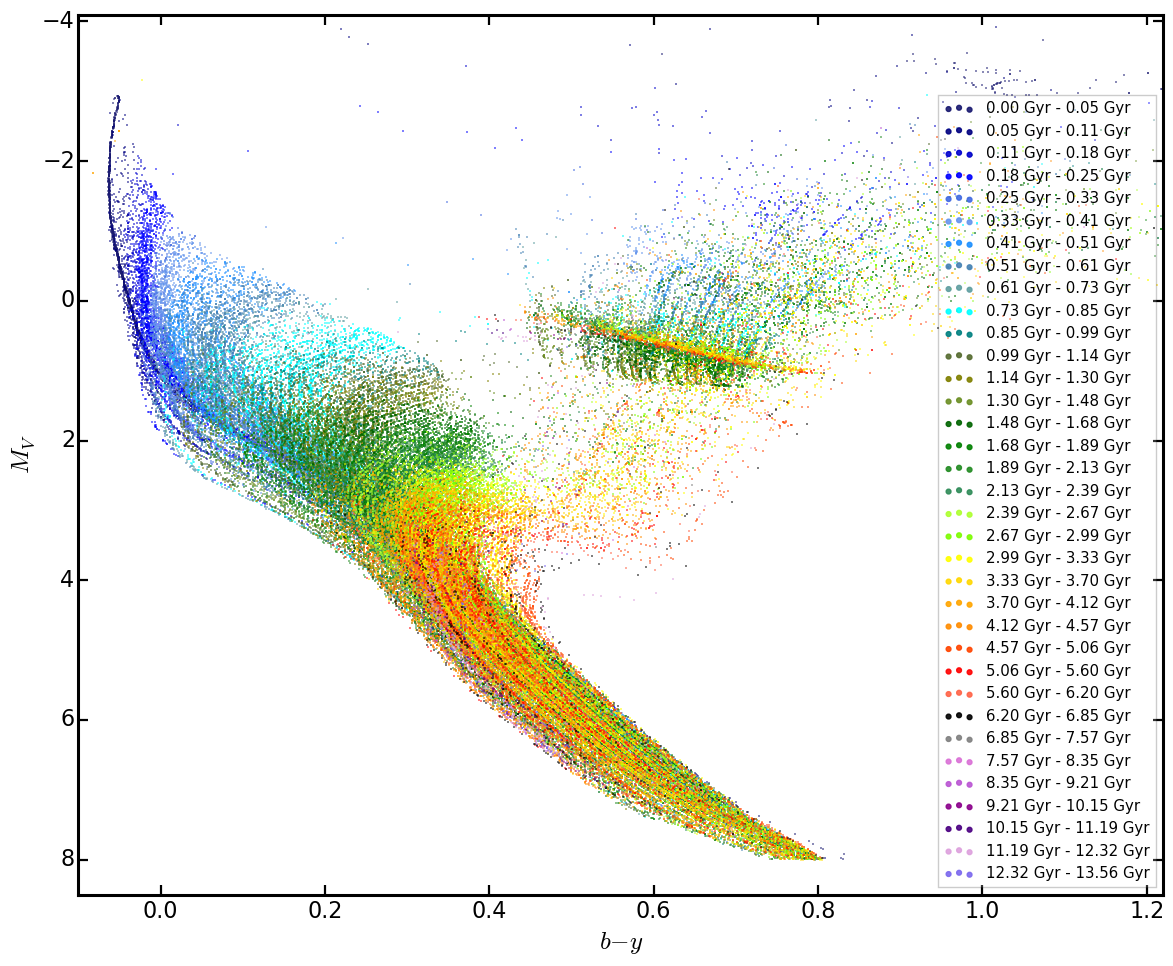} 
\end{minipage} 
\caption{Colour-absolute magnitude diagram for all the stars brighter than the apparent-magnitude limit 
	of the GCS ($m_V = 7.6$), i.e. the observational survey we want to compare our results with. 
	We show the absolute Johnson $M_V$ magnitude as a function of the 
	colour $b-y$, and consider stars evolving beyond the MS, too. Stars are colour-coded 
	according to their age, as highlighted in the legend.}
\label{CMassNoGcut} 
\end{figure*}

Uncertainties on metallicity estimates range between $0.04-0.05$ dex both in terms of overall metal content 
$[Z/H]$ and iron abundance $[Fe/H]$, and slightly depend on colours, increasing towards the blue-most and 
the red-most indices.
Stellar age determination is considered correct whenever either the absolute error is lower than $1$ Gyr or 
the relative uncertainty is higher than $0.25$.
Therefore, uncertainties on stellar ages are usually larger for older objects, the aformentioned criteria 
leading to good determination of absolute ages for young objects and to fair assessment of 
relative age determination for older stars.

The majority of stars in the GCS were unaffected by the extinction due to interstellar dust, 
these stars being located in the solar neighbourhood. 
A reddening correction is assumed for stars with a color excess $E(b-y) \ge 0.01$ mag and 
farther than $40$ pc from the observer, otherwise no correction was applied. Only $\sim 25 \%$ of the sample 
needed such a reddening correction, the mean reddening being $0.0025$ mag 
\citep[see][for further details]{Holmberg2007, Casagrande2011}.

As for the completeness of the survey, the GCS is volume-complete to a distance of $40$ pc. 
While the commonly quoted apparent-magnitude limit of the sample is $m_V = 8.3$, 
the magnitude at which incompleteness sets depends on colour.
The CGS begins to lose completeness near an apparent-magnitude $m_V = 7.6$ and below $1$ M$_{\odot}$.

\section{Results}
\label{Results}

In this section we present our results. In Section \ref{theHRplot} we show the CMDs that we can obtain by 
applying the approach we have developed to the output of a cosmological simulation of a disc galaxy. 
In Section \ref{otherResults} we investigate several properties of our catalogue of synthetic stars and address 
the comparison with observations. 

Throughout Sections \ref{theHRplot} and \ref{otherResults} we focus on our reference simulation AqC5--fid, 
and in particular on the sub-volume across its galactic disc 
(centred at $r_{\rm Sun} = 8.33$ kpc and $\vartheta_{\rm Sun}=60 ^{\circ}$) 
that we have selected to analyse a solar neighbourhood-like region (see Section \ref{solarNeigh}). 
Simulations AqC5--cone and AqC5--3sIMF will be considered for comparison when 
we analyse the metallicity distribution function (Figure \ref{MDFtot}).

\begin{figure*}
\newcommand{\captionfonts}{\small}
\begin{minipage}{\linewidth}
\centering
\includegraphics[trim=0.0cm 0.0cm 0.0cm 0.2cm, clip, width=0.785\textwidth]{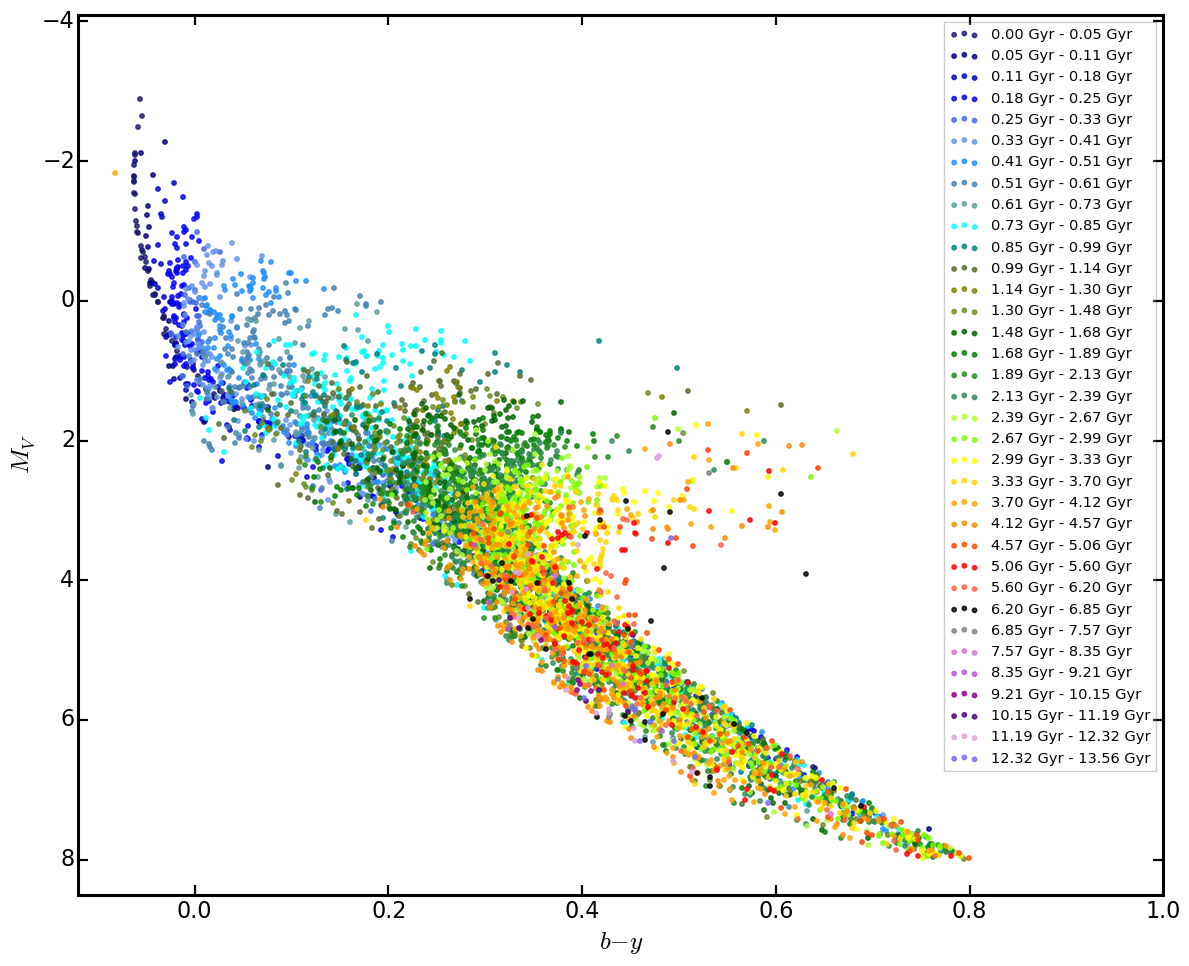} 
\includegraphics[trim=0.0cm 0.2cm 0.0cm 0.2cm, clip, width=0.795\textwidth]{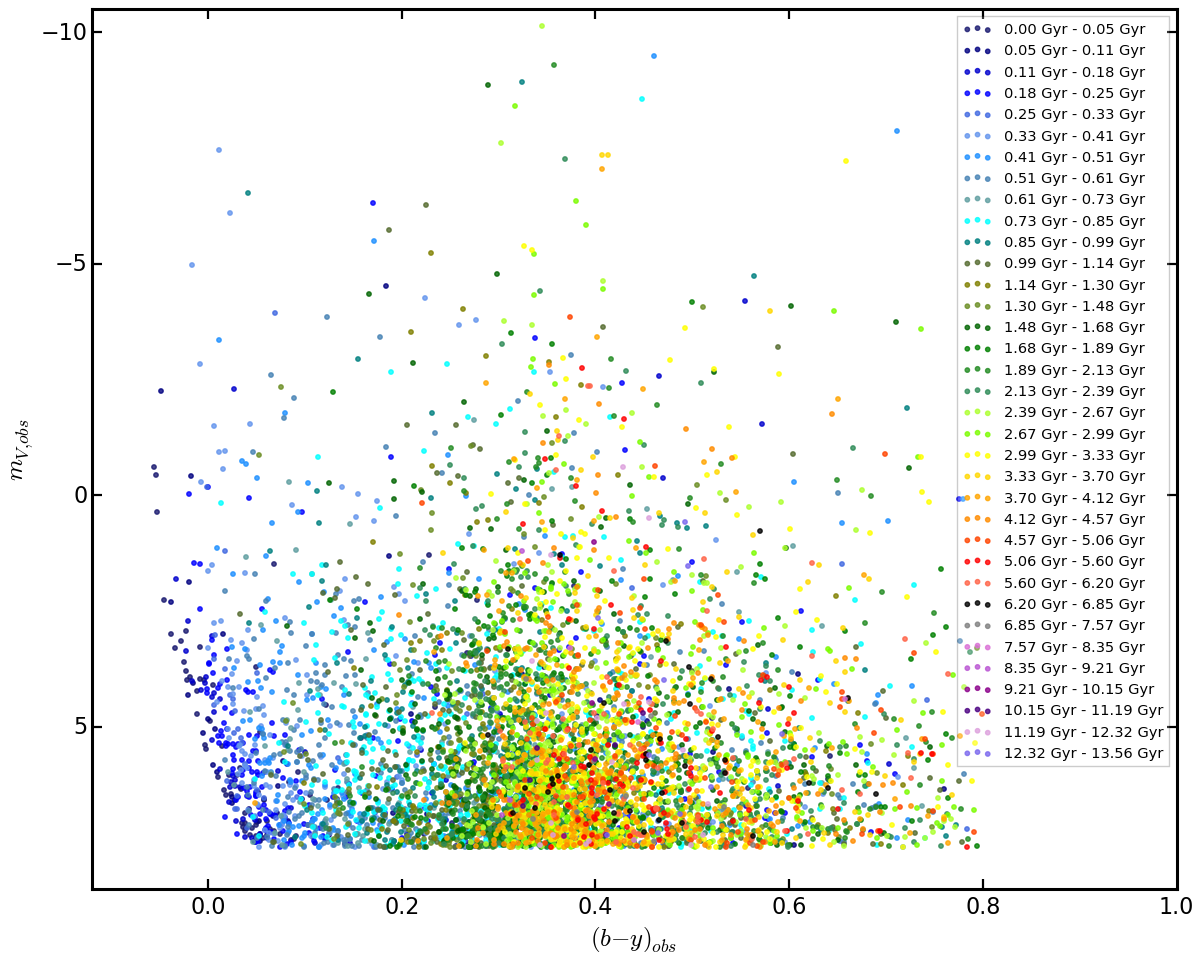} 
\end{minipage} 
\caption{Colour-absolute magnitude diagram (top panel) and 
	CMD (i.e. colour-apparent magnitude diagram, bottom panel) for the sample ($6670$ stars) that matches 
	the observational data set of the GCS. 
	Stars are colour-coded according to their age, as highlighted in the legend.
	These stars are brighter than the apparent-magnitude limit of the survey ($m_V = 7.6$) and have a surface 
	gravity typical of MS stars (log $g \ge 3.0$). In the top panel we show the absolute Johnson $M_V$ magnitude 
	as a function of the colour $b-y$. In the bottom panel we consider the apparent Johnson $m_{\rm V, obs}$
	magnitude (accounting for dust extinction, too) as a function of the reddened colour $(b-y)_{\rm obs}$.}
\label{CMappAss6670} 
\end{figure*}

\subsection{CMD from cosmological simulations}
\label{theHRplot}

We start our analysis by investigating properties of mock stars in the aformentioned selected region of AqC5--fid.
Figure \ref{CMassNoGcut} introduces the colour-absolute magnitude diagram. We show the absolute 
Johnson $M_V$ magnitude as a function of the colour $b-y$ for all the stars in our catalogue that are brighter 
than the apparent-magnitude limit of the survey. We consider $m_V = 7.6$ for the GCS apparent-magnitude limit. 
Here, we do not restrict the sample of stars to those whose surface gravity is larger than log $g \ge 3.0$, 
so as to appreciate all the possible evolutionary stages and not to limit to MS stars alone. 
Stars are colour-coded according to their age, that is the age of their own parent particles. 
The various filters and bands in which magnitudes are considered reflect the ones adopted in the GCS 
\citep[see][]{Casagrande2011}. 
Stars sharing the same colour can be deemed as an ensemble of stars lying on a set of isochrones characterised 
by similar ages (ranging between each age bin). Isochrones are curves on the luminosity 
(or absolute magnitude)-temperature (or colour) plane that link evolutionary stages of stars with different mass 
at a given time. Each isochrone is identified by age and chemical composition. In Figure \ref{CMassNoGcut} we 
can appreciate the spread in metallicity of quasi-coeval stars by analysing how stars of the same age that are 
experiencing the MS stage extend over the colour index. 
Assuming the same age, stars with a higher metal content have both a larger colour index and a larger mass at 
the turn-off (TO) point. Also, the TO becomes fainter and redder (higher $b-y$) as the matallicity increases. 
Above the TO, the extent across the colour index of coeval stars is mainly due to the different evolution 
that stars with different mass undergo.
Figure \ref{CMassNoGcut} shows how stars of different colours describe different shapes: as time passes, 
stars of different mass indeed evolve out of the MS, thus revealing striking features of aging stellar populations. 
For instance, the TO progressively dims and reddens as the stellar age increases; the red giant branch (RGB) 
becomes well populated and more extended for stars older than $\sim 1$ Gyr; also, the luminosity of the 
horizontal branch (HB) is roughly constant ($M_V \sim -1 \div 0$) after $\sim 1$ Gyr. 
Stars characterised by the lowest values of $M_V$ show an interesting feature: their $M_V$ progressively 
decreases as their colour index decreases. The reason for this behaviour stems from the threshold mass of 
our sample ($0.6$ M$_{\odot}$, see Section \ref{Generation of mock stars} and Figure \ref{MassMet}): stars 
having the same mass but lower metallicity (i.e. left edge of the MS wideness) have a higher temperature and 
a lower $M_V$, due to their lower opacity.

We then analyse properties of the $6670$ stars that make up the final stellar catalogue, i.e. the one that matches 
the observational data set of the GCS (see Section \ref{Data}).
Figure \ref{CMappAss6670} shows the colour-absolute magnitude diagram (top panel) and 
the CMD (colour-apparent magnitude diagram, bottom panel) for the sample of $6670$ stars. 
Stars are colour-coded according to their age, as described in the legend. We adopted $m_V = 7.6$ as  
GCS apparent-magnitude limit and log $g \ge 3.0$ as threshold surface gravity. 
In the top panel of Figure \ref{CMappAss6670} we consider the absolute Johnson $M_V$ magnitude as a function 
of the colour $b-y$: we can appreciate the effect of the surface gravity limit in restricting the sample to 
unevolved stars alone by contrasting with Figure \ref{CMassNoGcut}, where the low-magnitude portion of the 
plane is well populated for $b-y \gtrsim 0.4$. 
The bottom panel of Figure \ref{CMappAss6670} illustrates the apparent CMD of our star sample. Here, the 
Johnson $m_{\rm V, obs}$ magnitude has been corrected for the effect of dust (see equation (\ref{dust}), 
Section \ref{CMD}) and the color index $(b-y)_{\rm obs}$ is reddened according to equation (\ref{reddening}). 
The effect of the apparent-magnitude limit is evident here, since the diagram is not populated 
below $m_{\rm V, obs} = 7.6$. The majority of stars locate where $m_{\rm V, obs} \lesssim 3$ and is characterised 
by a colour index $0 \lesssim b-y \lesssim 0.6$. The distribution of stars gets sparser as the magnitude decreases 
and as the colour index depicts lower temperatures. 
These CMDs show the observable stellar content of a volume of our simulated galaxy.

\subsection{Properties of synthetic stars}
\label{otherResults}

\begin{figure}
\newcommand{\captionfonts}{\small}
\hspace{-2.1ex}
\includegraphics[trim=0.1cm 0.0cm 0.0cm 0.0cm, clip, width=0.5\textwidth]{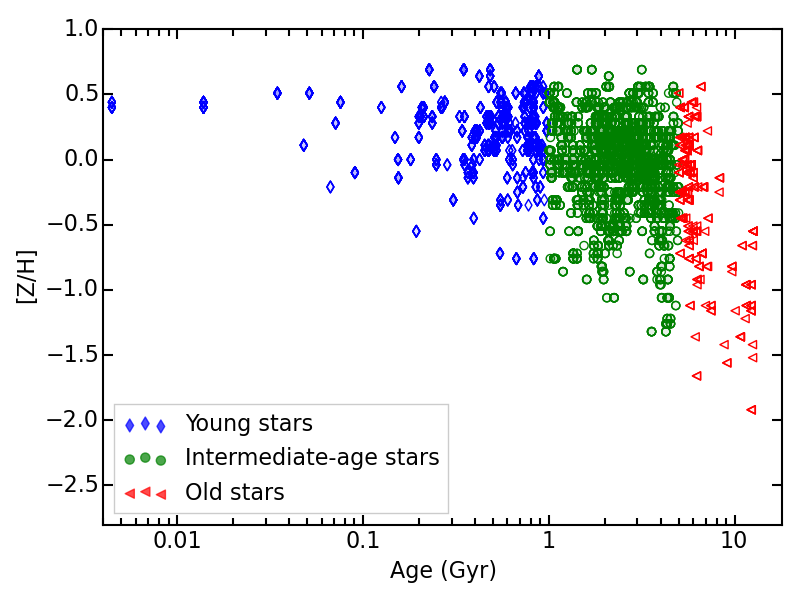} 
\caption{Age-metallicity distribution for the sample of $6670$ stars, divided into different age intervals. 
	Metallicity is analysed in terms of $[Z/H]$. Stars are colour-coded according to their age 
	(i.e. the age of their own parent particles, see text): blue diamonds show stars younger than $1$ Gyr, 
	green circles depict stars with age between $1$ and $5$ Gyr, while stars older than $5$ Gyr 
	are identified by red triangles. Stars share age and a similar metallicity with their parent star particles: 
	darker-filled symbols with thicker edges identify stars coming from more populated parent particles. }
\label{AgeMetStars} 
\end{figure}

We continue our analysis by further investigating properties of the catalogue of $6670$ mock stars.
Figure \ref{AgeMetStars} illustrates the AMD for the sample of $6670$ stars. Stars are colour-coded 
according to their age.
The identifying colour convention is the same adopted for parent particles in Figure \ref{Age-metallicityRelation}. 
Metallicity is here analysed in terms of $[Z/H]$.
Stars have been generated according to the input properties of the selected parent particles: 
therefore, stars have the same age and a similar metallicity of their own parent particles. In particular, 
the $[Z/H]_{\rm stars}$ ranges between $[Z/H]_{\rm parent \, particle} \pm 0.05$ dex (see Section \ref{CMD}). 
This explains why stars and star particles share an almost indistinguishable AMD 
(and proves the accuracy of the stochastic sampling, too). 
In Figure \ref{AgeMetStars}, the shade of each symbol is proportional to the number of stars sharing 
properties on the age-$[Z/H]$ plane: 
thicker symbols identify stars coming from more sampled parent particles 
(see equation (\ref{Nextr})). 
The reason why the AMDs of parent particles and stars are 
not exactly the same is twofold: first, stars associated to the same star particle can have sligthly different 
$[Z/H]$ (see, for instance, the youngest stars on the left of Figure \ref{AgeMetStars}). This is due to the 
fact that stars in two contiguous metallicity bins can have an $[Z/H]_{\rm stars}$ within 
$[Z/H]_{\rm parent \, particle} \pm 0.05$ dex. Also, some parent particles 
can be not represented at all by the final $6670$ stars, either because they are not massive enough to generate 
stars that pass the selection and extraction procedures, or because their child stars have an apparent magnitude 
that is higher than the apparent-magnitude limit of the GCS (for instance, by comparing Figures 
\ref{Age-metallicityRelation} and \ref{AgeMetStars}, we see that the star particles with a 
$[Z/H]_{\rm parent \, particle} < -2$ are not represented by stars).

\begin{figure}
\newcommand{\captionfonts}{\small}
\hspace{-2.1ex}
\includegraphics[trim=0.1cm 0.0cm 0.0cm 0.0cm, clip, width=0.5\textwidth]{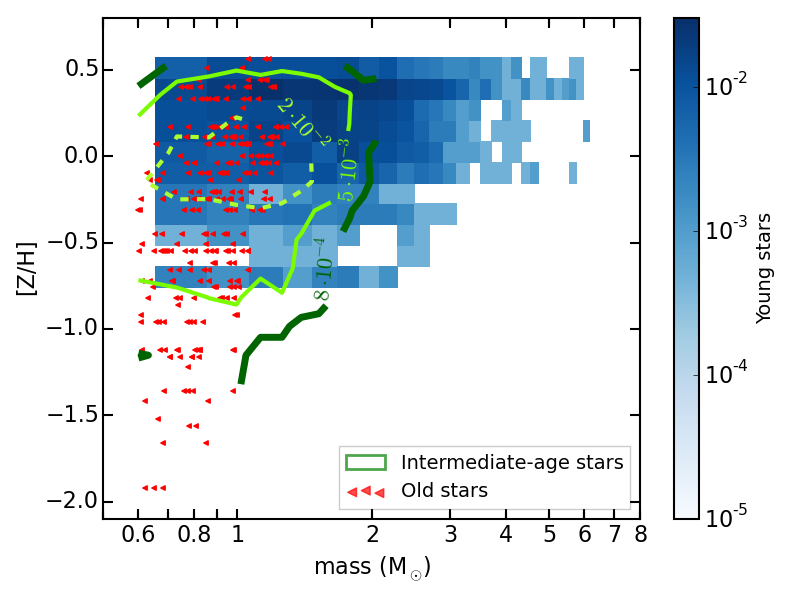} 
\caption{Mass-metallicity distribution for the $6670$ star catalogue, once it has been split into three subsamples 		according 
	to stellar age. The background two-dimensional blue histogram represents the distribution of young stars, the 
	bins being colour-coded according to the fraction of stars with respect to the total number of young stars. Green 
	contours describe the distribution of intermediate stars: from outwards to inwards contours enclose bins 
	with a fraction of $8 \cdot 10^{-4}$, $5 \cdot 10^{-3}$, and $2 \cdot 10^{-2}$ stars with respect to the total number 
	of intermediate stars, respectively. Red triangles show the distribution of old stars.}
\label{MassMet} 
\end{figure}

Figure \ref{MassMet} shows the mass-metallicity distribution for our star sample. 
The $6670$ star sample has been 
sliced into three subsamples -- young, intermediate-age, and old stars -- according to the aformentioned age 
classification. The young (blue), intermediate-age (green) and old (red) star subsamples consist of the following 
number of stars: 
$2047$, $4351$, and $272$, respectively. 
The background two-dimensional histogram represents the distribution of young stars in this plane. 
Mass-metallicity bins are colour-coded according to the fraction of stars in the bin with respect to the total number 
of young stars, the fraction of stars increasing from ligther- to darker-blue bins. 
Green contours show the distribution of intermediate stars: from outwards to inwards contours enclose 
mass-metallicity bins with a fraction of $8 \cdot 10^{-4}$, $5 \cdot 10^{-3}$, and $2 \cdot 10^{-2}$ stars with respect to 
the total number of intermediate-age stars, respectively. Red triangles describe the distribution of old stars. 
This figure displays the mass distribution of stars that make up our sample: the majority of stars have a 
mass ranging between $0.6$~M$_{\odot}$ and $\sim 1.5$~M$_{\odot}$, with mass never exceeding $7$~M$_{\odot}$. 
Mass distribution is more clustered going from young to intermediate-age to old stars, only few old stars 
having a mass larger than $1$~M$_{\odot}$. 
Young stars have on average a higher $[Z/H]$, although the region of the plane with $-0.1 \lesssim [Z/H] \lesssim 0.3$ 
is almost equally populated by young and intermediate-age stars (note that the blue and green subsamples differ  
by a factor of $\sim 2$ in terms of number of stars, so numbers expressing fractions almost immediately translate 
in number of stars). 
As the stellar age increases, stars tend to move towards smaller masses 
(as a consequence of stellar evolution) and lower metallicities. 
As a consequence, higher-metallicity stars spread over the entire mass range, while the lower the $[Z/H]$ is, the 
smaller is the mass of the star and the older is the star. 

\begin{figure}
\newcommand{\captionfonts}{\small}
\hspace{-2.1ex}
\includegraphics[trim=0.1cm 0.0cm 0.0cm 0.0cm, clip, width=0.5\textwidth]{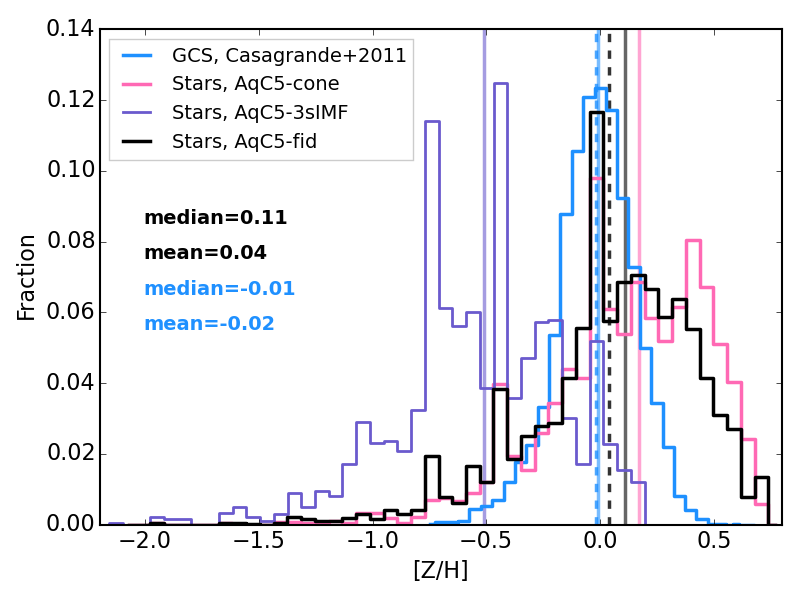} 
\caption{MDF for the star sample of $6670$ stars in AqC5--fid (black histogram), in AqC5--cone (pink histogram), 
	and AqC5--3sIMF (violet histogram). Solid (and dashed) vertical lines 
	indicate the median (and the mean) of each distribution. We also show the MDF 
	obtained for the GCS by \citet{Casagrande2011}, for comparison (light blue histogram). Solid and dashed 
	light blue vertical lines describe the median and the mean of the latter distribution, respectively. 
	}
\label{MDFtot} 
\end{figure}

Figure \ref{MDFtot} introduces the metallicity distribution function (MDF) for models and observation. 
We consider here three different star samples, one for each simulation that we carried out 
(see Section \ref{TheSim} and Table \ref{TableParamm}). 
The three star samples represent the observable stellar content of three volumes, 
each centred at $r_{\rm Sun} = 8.33$ kpc and $\vartheta_{\rm Sun}=60 ^{\circ}$ 
in the reference simulation AqC5--fid, in AqC5--cone and in AqC5--3sIMF. 
The black histogram in Figure \ref{MDFtot} depicts the MDF of the star catalogue of AqC5--fid, 
violet and pink histograms describing the MDFs of the star samples of AqC5--3sIMF and AqC5--cone, respectively. 
We also show the MDF obtained by \citet{Casagrande2011} for the GCS (in light blue), 
and we report their median and mean values ($[Z/H]= -0.01$, $[Z/H]= -0.02$, respectively).

\begin{figure}
\newcommand{\captionfonts}{\small}
\hspace{-2.1ex}
\includegraphics[trim=0.1cm 0.0cm 0.0cm 0.0cm, clip, width=0.5\textwidth]{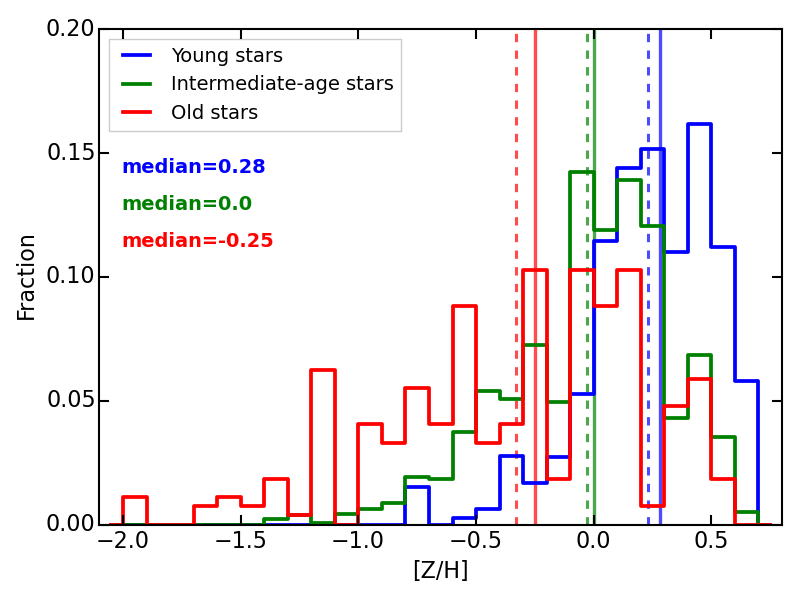} 
\caption{MDF for the three subsamples of stars in which the $6670$ star catalogue has been sliced 
	according to stellar age. Solid and dashed vertical lines pinpoint the median 
	and the mean of each distribution, respectively. The $6670$ star catalogue is made up of $2047$ young stars, 
	$4351$ stars with age between $1$ and $5$ Gyr, and $272$ old stars.}
\label{MDFsplit} 
\end{figure}

\begin{figure}
\newcommand{\captionfonts}{\small}
\hspace{-2.1ex}
\includegraphics[trim=0.1cm 0.0cm 0.0cm 0.0cm, clip, width=0.5\textwidth]{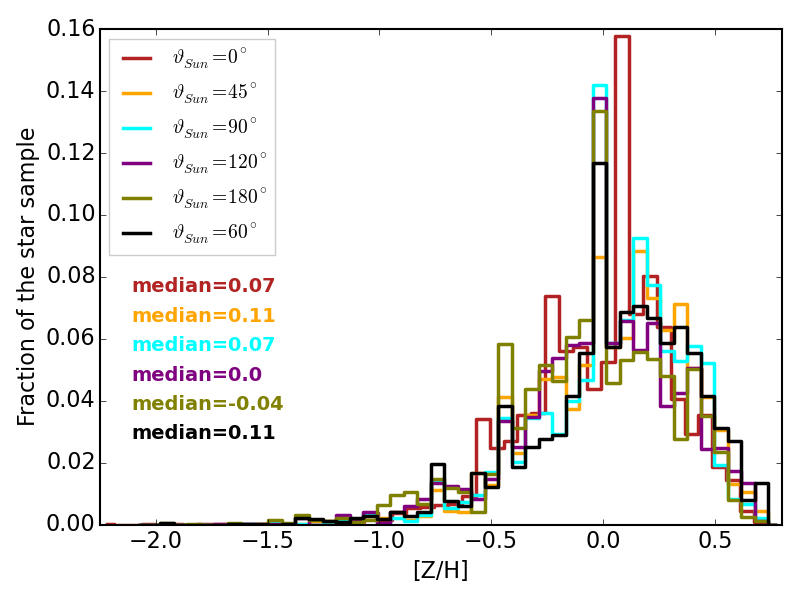} 
\caption{MDFs for six different star samples of $6670$ stars each for the AqC5--fid galaxy. 
	These star samples are located in volumes whose origin is at fixed distance from the 
	galaxy centre, i.e. $r_{\rm Sun} = 8.33$ kpc, whereas their position angle varies on the galactic plane 
	of the simulated galaxy. We consider $\vartheta_{\rm Sun} = 0^{\circ}$, $\vartheta_{\rm Sun} = 45^{\circ}$, 
	$\vartheta_{\rm Sun} = 60^{\circ}$, $\vartheta_{\rm Sun} = 90^{\circ}$, $\vartheta_{\rm Sun} = 120^{\circ}$, 
	and $\vartheta_{\rm Sun} = 180^{\circ}$, respectively. We also indicate the median 
	of each distribution. Note that the black MDF identifies the same distribution shown in 
	Fig. \ref{MDFtot} and Fig. \ref{DiffRadii}.}
\label{DiffAngles} 
\end{figure}

\begin{figure}
\newcommand{\captionfonts}{\small}
\hspace{-2.1ex}
\includegraphics[trim=0.1cm 0.0cm 0.0cm 0.0cm, clip, width=0.5\textwidth]{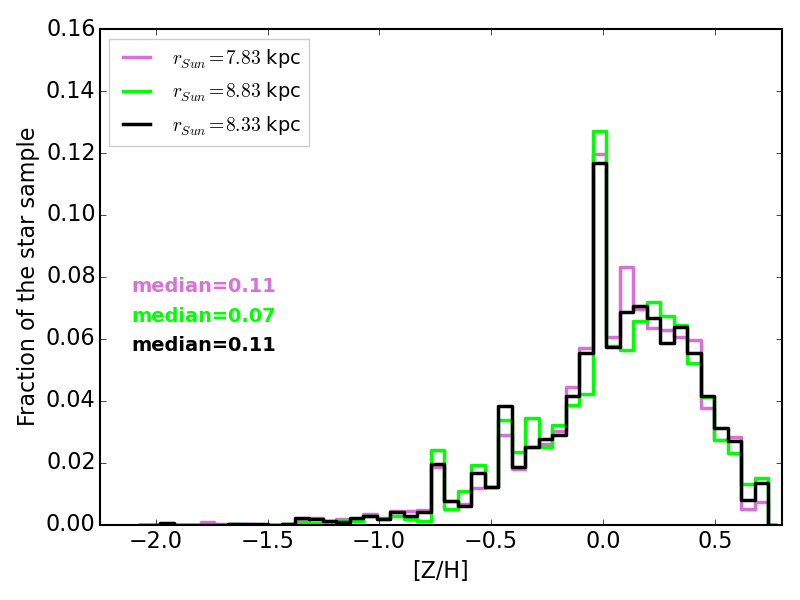} 
\caption{MDFs for three different star samples of $6670$ stars each for the AqC5--fid galaxy. 
	These star samples are located in spheres whose origin is at increasing distance from the 
	galaxy centre, i.e. $r_{\rm Sun} = 7.83$ kpc, $8.33$ kpc, and $8.83$ kpc, respectively. The centre of 
	all these volumes is on the galactic plane of the simulated galaxy, at a fixed position angle 
	($\vartheta_{\rm Sun} = 60^{\circ}$). Note that the black MDF identifies the same distribution shown 
	in Fig. \ref{MDFtot} and \ref{DiffAngles}.}
\label{DiffRadii} 
\end{figure}

Before drawing any strong conclusion from this comparison, we caution that the metallicity scale in 
observations and simulations could be different. The former is derived from photometry, 
the latter is inferred directly from our chemical evolution network 
and then accounts for the binning discussed in Section \ref{Generation of mock stars}. This caveat has to 
be considered when comparing our models with observations. For this reason, we will focus on 
the overall shape and mean values of predicted and observed MDFs. We choose a bin size 
that roughly matches that of the histogram in \citet{Casagrande2011} for the MDF of our models. 
The MDF of the solar neighbourhood-like volume in AqC5--fid (median $[Z/H]= 0.11$, mean $[Z/H]= 0.04$) 
agrees well with observations. The means and medians of the two distributions differ by only $0.06$ dex 
and $0.12$ dex, respectively (see also Figures \ref{DiffAngles} and \ref{DiffRadii}, and Section \ref{Dipendenze}). 
The difference between the means of the two MDFs implies that stars of our sample 
are on average richer in metals than observed ones by $\sim 15 \%$. 
The considered star sample of AqC5--fid is characterised by a 
low-metallicity tail that is more extended than the MDF of the GCS, and that is mainly populated by old stars 
(see Figure \ref{MDFsplit}). Also, a larger fraction of mock stars has a super-solar metallicity with respect to 
the observed sample. 
Despite these differences, the peaks of the two MDFs perfectly overlap and pinpoint a solar metallicity.

The peak at solar metallicity in the MDF of AqC5--fid is not within the Poisson errors computed on the number 
of mock stars in metallicity bins 
(this is true for the main peaks in the MDFs of AqC5--cone and AqC5--3sIMF, too). By fitting the MDF of AqC5-fid 
with a Gaussian (mean $0.117$, standard deviation $0.317$) we find that this peak is $11.7 \sigma$ above 
the value predicted by the fit ($\sigma \sim 0.004$ being the Poisson error of the considered bin).
The peak in the MDF of mock stars is due to the properties of the parent particles closer to the centre of the solar neighbourhood-like volume. These parent particles generate mock stars that are more likely to pass the apparent magnitude limit cut. The metallicity of the peak therefore traces the metallicity of the innermost region of the galactic 
disc. 

When associating mock stars to their parent particles, we take a possible uncertainty in metallicity into account 
(i.e. $0.05$ dex, see Section \ref{CMD}). This uncertainty in metallicity means that we 
distribute the same mock stars to star particles located into two contiguous metallicity bins (see above in this section). 
However, it can happen that some parent particles are isolated in the age-metallicity plane, without other particles 
populating metallicity bins directly above or below. For parent particles located into these isolated bins, mock stars 
will be associated with a precise value of metallicity. This means that the uncertainty on metallicity estimates 
for mock stars whose metallicity is that of these isolated bins will be underestimated with respect to observations. 
This is reflected on the fact that when we derive the MDF for our mock stars with the same binning as in observations 
we are recovering into a single bin mock stars that should populate also adjacent metallicity bins.
This can be solved by introducing a slight smoothing in our distribution. If we add to the metallicity of each mock star 
an uncertainty drawn from a normal distribution (centred on the metallicity of the considered mock star and with 
standard deviation $0.025$ dex), the peaks of predicted MDFs in Figure \ref{MDFtot} are considerably reduced. 

We also consider the MDFs of synthetic star catalogues in AqC5--cone and AqC5--3sIMF.
Stars in the selected sphere of AqC5--cone have a MDF with slightly super-solar median and mean values 
($[Z/H]= 0.17$, $[Z/H]= 0.10$, respectively). The excess of metal-rich stars is a consequence of the 
low-redshift SFR (higher than in the other galaxies, see Figure \ref{sfr}), that affects outer regions of the galaxy disc, 
further enriching them in metals. The higher metal content of stars in AqC5--cone with respect to AqC5--fid 
is also supported by the radial metallicity profiles of these two galaxies shown in \citet{Valentini2017}. 
As for AqC5--3sIMF, the peak of the MDF is subsolar (median $[Z/H]= -0.51$, mean $[Z/H]= -0.53$). 
The distribution is left-skewed and is characterised by a pronounced low-metallicity tail. 
The comparison with the GCS suggests that stars in the solar neighbourhood are metal richer than 
those in AqC5--3sIMF by a factor of $\sim 2.5$ (the medians of the two distributions differ by $0.4$ dex). 
The tension between predictions from AqC5--3sIMF and the reference AqC5--fid is due to the different IMF 
(see Section \ref{TheSim}): the K3s produces a smaller number of massive stars than K2s, thus limiting 
the metal enrichment contributed by SNe~II (the role of different IMFs will be thoroughly addressed in 
Valentini et al., in preparation).
However, we note that \citet{Casagrande2011} also cautioned that it is not straightforward to compare 
directly their MDF with theoretical predictions, since it is affected by sample selection effects. For instance, 
they provide the MDF for a reduced number of stars ($5976$) with respect to their $irfm$ sample. 
They indeed consider for the MDF analysis only those stars that meet the criteria needed for their metallicity 
calibration. While doing that, they exclude some of the stars whose colour index is in the range 
$0.43 \le (b-y) \le 0.63$: it is not trivial to figure out how the aforementioned sample selection effect 
affects the MDF of observed stars. 
In Section \ref{Dipendenze} we discuss how the shape and the peak of the MDF are affected by the position 
in the simulated galaxy of the volume where we carry out our analysis.

We then consider how the three subsamples of young, intermediate-age, and old stars in which we split the 
whole star catalogue contribute to the total MDF. Here we focus our analysis on stars within the volume 
centred at $r_{\rm Sun} = 8.33$ kpc and $\vartheta_{\rm Sun}=60 ^{\circ}$ in the reference simulation AqC5--fid.
Figure \ref{MDFsplit} shows the MDF for the three subsamples. The age classification is the same adopted in 
Figures \ref{Age-metallicityRelation}, \ref{AgeMetStars}, and \ref{MassMet}, and discussed above in this section. 
The reason why in Figure \ref{MDFsplit} we prefer to show split MDFs in terms of fraction of stars with given 
metallicity relative to each entire subsample lies in the uneven contribution by different subsamples. 
The old star MDF (mean $[Z/H] = -0.33$, median $[Z/H] = -0.25$) shows that the low-metallicity tail of 
the total MDF of AqC5--fid in Figure \ref{MDFtot} is mainly contributed by old stars. 
As already pointed out by the mass-metallicity distribution in Figure \ref{MassMet}, 
young stars are characterised by 
the narrowest distribution; the MDF of intermediate-age stars (mean $[Z/H] = -0.03$, median $[Z/H] = 0.00$) 
has a broader barely-populated wing that extends towards very low $[Z/H]$. 
Moreover, the blue MDF (mean $[Z/H] = 0.23$, median $[Z/H] = 0.28$) shows that young stars are the main 
responsible for the quasi-totality of stars with highly super-solar metallicity, i.e. $[Z/H] > 0.3$.  
Also, young and intermediate-age stars are the main contributors to stars with the
$[Z/H]$ value where the total MDF peaks (see Figure \ref{MDFtot}).

\section{Discussion}
\label{Discussion}

\subsection{Solar neighbourhood sampling variance}
\label{Dipendenze}

In this section we address the impact of the assumed position of the solar neighbourhood in the simulated 
galaxy on final results. Do our conclusions change if we choose different positions for the centre of the 
volume in which we perform our analysis? In Section \ref{TheSim} we stressed that our simulated disc galaxies 
should not be considered as a model of the MW, in spite of similar morphological properties and several features 
shared by them. Also, when we selected the sphere where our analysis has been carried out 
(Section \ref{solarNeigh}), we did not make any particular attempt to reproduce the actual location of 
the solar neighbourhood with respect to the bar orientation and the spiral pattern of the MW, nor we 
considered the exact radial extent or the scale radius of the stellar disc of the simulated galaxies. 
However, we compare our predictions with results from a typical survey in the solar neighbourhood: 
thus, we assume that the variation of properties of star samples at marginally different positions is not 
dramatic once a star sample large enough to be representative of a typical region of the galactic disc is chosen.

In order to investigate if sample selection effects modify final results, we select different spheres in the AqC5--fid 
galaxy, we extract from each a star catalogue made up of $6670$ stars (as we did for our reference case 
in previous sections), and we compare MDFs of different star samples. All the volumes where we 
perform the analysis have origins on the galactic plane: we first fix a distance $r_{\rm Sun}$ and consider 
different $\vartheta_{\rm Sun}$. Then, we rather choose a position angle $\vartheta_{\rm Sun}$ for their centres 
and allow their distance from the galaxy centre to vary by $0.5$ kpc from time to time. In all the 
considered cases, we keep the size of the sphere constant (i.e. $1$ kpc).

Figure \ref{DiffAngles} highlights that MDFs of stars located on the galactic plane of our simulated galaxy are 
comparable, whatever is the angle $\vartheta_{\rm Sun}$ that identifies the location of the centre of the 
volume in which we perform the analysis.
We introduce the MDFs for six different star samples. These star samples are all located in spheres 
whose origin is $r_{\rm Sun} = 8.33$ kpc far from the galaxy centre. 
On the other hand, their position angle varies on the galactic plane of the simulated galaxy. 
We consider $\vartheta_{\rm Sun} = 0^{\circ}$, $\vartheta_{\rm Sun} = 45^{\circ}$, 
$\vartheta_{\rm Sun} = 60^{\circ}$, $\vartheta_{\rm Sun} = 90^{\circ}$, $\vartheta_{\rm Sun} = 120^{\circ}$, and
$\vartheta_{\rm Sun} = 180^{\circ}$, respectively. 
The median of each distribution is as follows: 
$0.07$, $0.11$, $0.11$, $0.07$, $0.00$, and $-0.04$, respectively, the medians of 
some MDFs overlapping.
Note that the black MDF identifies the same distribution shown in Figures \ref{MDFtot} and \ref{DiffRadii}. 
Also, we notice that in Section \ref{otherResults}, Figure \ref{MDFtot} we have analised the MDF for the star sample 
in the reference volume ($r_{\rm Sun} = 8.33$ kpc, $\vartheta_{\rm Sun} = 60^{\circ}$): that MDF is characterised 
by a median value that is among the ones that differ the most by the median of the observed MDF in the GCS, 
and this highlights the agreement of our results with observations. 
All the MDFs in Figure \ref{DiffAngles} peak around solar metallicities, the median values of different distributions 
differing by $0.15$ dex at most. 
We can therefore conclude that sample selection effects do not affect significantly our results.
We note that the result shown in Figure \ref{DiffAngles} also ensures that differences in metal content among 
star particles (and thus among stars) within regions whose radius is $1$~kpc are trustworthy.
Since our formal resolution is $\sim 1$ kpc, we are relying on a single sampling of the interested volume 
when we analyse stellar properties in one sphere. However, retrieving comparable MDFs within different volumes 
shows that statistical fluctuations cannot be responsible for the analysed chemical features.

\begin{figure*}
\newcommand{\captionfonts}{\small}
\begin{minipage}{\linewidth}
\centering
\includegraphics[trim=0.4cm 0.0cm 1.15cm 0.0cm, clip, width=0.497\textwidth]{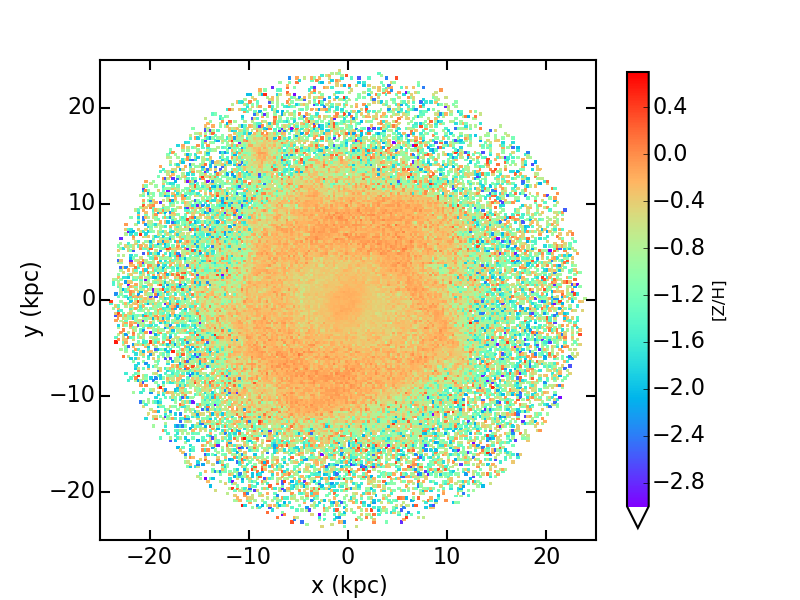} 
\includegraphics[trim=0.4cm 0.0cm 1.15cm 0.0cm, clip, width=0.497\textwidth]{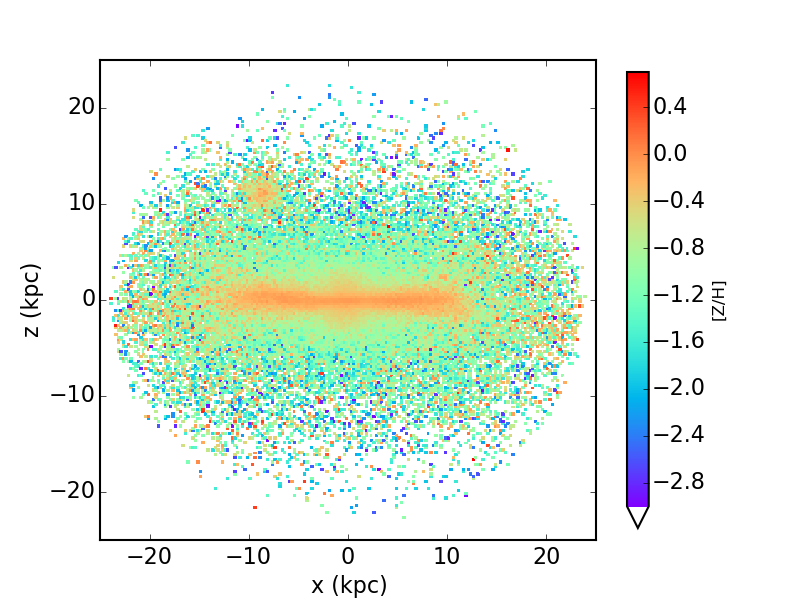} 
\end{minipage} 
\caption{Face-on (left-hand panel) and edge-on (right-hand panel) binned distributions of all the star particles located 
within the galactic radius for the AqC5--fid galaxy simulation. Plots are shown at redshift $z = 0$; the colour encodes 
the mean $[Z/H]$ of the star particles in the bin.}
\label{MetDensityMaps} 
\end{figure*}

Figure \ref{DiffRadii} shows the MDFs for three different star samples of $6670$ stars each. The volumes 
where these star samples are located are centred at  
$r_{\rm Sun} = 7.83$ kpc, $r_{\rm Sun} = 8.33$ kpc, and $r_{\rm Sun} = 8.83$ kpc, respectively. 
The centre of all these spheres is on the galactic plane of the simulated galaxy, at a fixed position angle, 
i.e. $\vartheta_{\rm Sun} = 60^{\circ}$. 
Each distribution is characterised by the following median: $0.11$, $0.11$, and $0.07$, respectively.  
Stars in selected spheres centred at increasing distance from the galaxy centre share very similar MDFs. 
As already discussed, the reason for similar values of metallicity for the peak of the MDFs stems from 
the properties of parent particles located within the innermost regions of the solar neighbourhood-like volume, 
i.e. on the inner galactic disc. 
The stellar chemical homogeneity shared by the analysed regions of the stellar disc can be understood 
by analysing the metallicity maps in Figure \ref{MetDensityMaps}. We show face-on and edge-one distribution 
of all the star particles located within the galactic radius $R_{\rm gal}$ of the AqC5--fid galaxy simulation, at 
redshift $z = 0$. The colour encodes the mean $[Z/H]$ of the star particles in each spatial bin: we can 
appreciate how the metal content in stars changes as a function of the distance from the galaxy centre across the 
galactic plane (left-hand panel of Figure \ref{MetDensityMaps}) and as a function of the height over it (right-hand panel). 
Stars on the galactic disc within a distance of $r \lesssim 10$ kpc from the galaxy centre have on average 
a solar metallicity, the mean $[Z/H]$ rapidly decreasing as one moves towards larger distances and higher 
galactic latitudes. 

Also, in \citet{Valentini2017} we have shown that radial metallicity profiles of all the stars are almost flat for 
$r \lesssim 10$ kpc (see Figure 11, last panel, for AqC5--cone, this feature being shared with AqC5--fid)\footnote{This 
	is also the reason why we decide not to evaluate a scaled position for the centre of our volume $r_{\rm Sun}$ 
	according to an effective radius of the simulated galaxy. For the sake of completeness, by fitting 
	the stellar surface density profile of AqC5--fid with an exponential law 
	$\Sigma(r) \propto {\text{exp}}(- r/r_s)$, $r_s$ being the scale radius, we find $r_s=3.25$ kpc. For instance, 
	\citet{mcMillan2011} found $r_s = 3.29 \pm 0.56$ kpc for the thick disc 
	($r_s= 3.00 \pm 0.22$ kpc for the thin disc) of MW.}; on the other 
hand, a radial metallicity gradient (for $r > 5$ kpc) is evident and in keeping with observations when only younger 
stars are considered (see Figure 17 of \citealt{Valentini2017}, right-hand panel, for both AqC5--fid and AqC5--cone).

Our general result is that properties of mock stars originated from star particles that are located in 
marginally different regions of the simulated galaxy are remarkably similar. We indeed find that stars 
at a given distance from the galaxy centre share a considerably comparable MDF, regardless of their exact 
position angle on the galactic plane. The stellar chemical homogeneity is preserved when 
considering stars located within a torus and when we marginally increase or decrease the distance of the 
volume where we carry out the analysis from the galaxy centre on the galactic plane.
Thus, our model predicts that chemical features of stars observed in the solar neighbourhood are likely shared 
by stars on the Galactic plane at a similar distance from the MW centre, and to some extent by stars beyond 
the conventional extension of the solar neighbourhood.

\subsection{Comparison with results from the GCS}
\label{ComparisonGCS}

We outlined in Section \ref{Data} the reasons that led us to choose the GCS for the comparison with 
observational data. We want here to stress 
that the comparison between our results and observational findings is not meant to validate our method. 
Our simulated galaxy is not indeed a model for the MW. Also, different results are found from different 
observational star catalogues, as we discuss in Section \ref{ComparisonOtherSurvey}, and marginally contrasting 
conclusions can be drawn when peculiar sample selection effects affect different data sets.
We rather selected one survey among the many others in order to show how the features of the observational 
sample one wants to compare with enter the procedure we developed. One could mimic observational selection 
effects so as to overcome biases intrinsic to a survey, but this is beyond the scope of the present work. 

There are several interesting features that our star sample shares with stars in the GCS composing the $irfm$ sample. 
The mass-metallicity distribution of both our study and the GCS is characterised by high-metallicity stars 
spreading throughout the complete mass range, whereas lower metallicity stars are mainly old stars and some 
intermediate-age stars. Making a direct comparison between values of metallicity is difficult, since we analyse the 
metal content of stars in terms of $[Z/H]$, while they consider both the mass- and age-metallicity distribution in 
terms of $[Fe/H]$. Translating one quantity into the other is not straightforward for stars of different age.

As for the AMD, we find that our star sample shares the same trend as in observations. Young and 
intermediate stars have, on average, a higher metal content than older ones. Also, the metallicity distribution 
progressively narrows when moving from old to younger stars. 

Albeit \citet{Casagrande2011} do not provide us with the exact number of stars that make up each of the three 
subsamples, by analysing the normalised age probability distribution for all the stars in their catalogue it is 
possible to see that the intermediate star sample is the one that by far contributes the most (see their Figure $13$). 
This feature is shared with our reference star catalogue, where stars having age in the range $1-5$ Gyr are $4389$ 
out of $6670$. Contributions to the catalogue from young and old stars follow, thus in keeping with the GCS data set.

We have already discussed the MDF of our star sample in comparison with the one of \citet{Casagrande2011} 
in Sections \ref{Results} and \ref{Dipendenze} (see Figure \ref{MDFtot}). 
Here we note that \citet{Nordstrom2004} found a mean of $[Z/H]= -0.14$ for the MDF of stars in the GCS, 
and a dispersion of $0.19$ dex around the mean: therefore, the mean of the MDF can vary by more than 
$0.1$ dex when different authors analyse slightly different subsamples of the same survey, 
by assuming their own calibrations for physical properties.
By further investigating the role of subsamples of different age in the MDF, we notice that we have a negligible 
contribution by old stars to the metal-rich wing of the MDF (see Figure \ref{MDFsplit}). 
The MDFs of young and intermediate-age stars roughly overlap in our sample for $-0.1 \le [Z/H] \le 0.3$. 
The distribution of intermediate-age stars extends toward lower metallicity down to $[Z/H] \sim -1.3$ and 
toward higher metallicity to $[Z/H] \sim 0.7$, whereas the metallicity of young stars covers the range 
$-0.5 \le [Z/H] \le 0.7$ and has a super-solar mean due to recent star formation.
Young stars in the GCS sample have a much narrower distribution than intermediate stars: still, the comparison 
is not trivial being the split MDF analysed only in term of $[Fe/H]$ in \citet{Casagrande2011}. 

When performing the analysis presented in this work, we did not tune any parameter of our cosmological 
hydrodynamical simulation in order to attempt to match observational findings. As for the simulation AqC5-3sIMF 
we stress here that, in contrast with the tension between the predicted MDF and the one observed in the GCS 
(see Figure \ref{MDFtot}), we find a very good agreement between observations and 
our results concerning both radial abundance gradients in the gas of our galaxy and different metal abundances 
and $\alpha$-enhancement in stars 
(these results will be presented in a forthcoming paper Valentini et al., in preparation).

\subsection{Comparison with the MDF of other surveys}
\label{ComparisonOtherSurvey}

Stars in the $irfm$ sample of the GCS have a comparable MDF in terms of $[Z/H]$ and $[Fe/H]$. 
The former is characterised by the following values for the mean, the median, and the full width at half maximum 
(FWHM): $-0.02$, $-0.01$, and $0.34$, respectively. 
The latter has a mean $[Fe/H] = -0.06$, a median value of $-0.05$, and the FWHM is $0.38$. 

Other authors have investigated the MDF of different samples of nearby stars in the MW. Results are usually 
shown in terms of $[Fe/H]$, being the oxygen abundance (oxygen is the most common metal and an accurate tracer 
of the total metallicity) difficult to retrieve \citep[see e.g.][]{Cayrel2004}.
For instance, \citet{Gilmore1995} investigated the $[Fe/H]$ distribution for a sample of stars located 
in the thick disc of MW, finding that it peaks at $-0.7$ dex. Studies by \citet[e.g.][]{Rocha-Pinto1996, Rocha-Pinto1998},  
\citet{Favata1997}, and \citet{Holmberg2007} agree on a $[Fe/H]$ distribution peaking around $-0.2 \div -0.3$.

A slightly sub-solar value for the mean of the MDF is supported by studies of e.g. 
\citet{Haywood2001} and \citet{Taylor2005}; for instance, \citet{Allende2004} found that luminous 
stars within $15$ pc from the Sun in their sample have a $[Fe/H]$ distribution with a mean $[Fe/H] = -0.11$, the
standard deviation being $\sim 0.2$ dex. 
On the other hand, investigations by e.g. \citet{Luck2006} confirm a marginally super-solar peak value of the MDF. 
Reasons why different results are found are claimed to be due to systematic errors, different calibrations 
especially in the effective temperature of stars, sample selection effects.

As for more recent results, the MDF of stars of the AMBRE Project \citep{deLaverny2013} located in the thin disc of 
the MW spans the range $-0.5 \lesssim [Fe/H] \lesssim 0.5$, the mean value being roughly solar; stars in this 
survey located in the thick disc have $-1 \lesssim [Fe/H] \lesssim -0.2$, and a $[Fe/H] \simeq -0.5$ mean value 
\citep[see][for instance]{Grisoni2017}. However, if stars with high metallicity and high $\alpha$-elements are 
included when considering stars in the thick disc (the AMBRE Project considers these components separately), 
the MDF peaks around a $[Fe/H]$ value that is slightly sub-solar.

\section{Summary and Conclusions}
\label{sec:conclusions}

In this work we have introduced a novel approach to link star particles in simulations with observed stars. 
The key goal of this new method is to compare as fairly as possible the outcome of cosmological simulations, 
that provide a coarse sampling of stellar populations by means of star particles corresponding to simple stellar 
populations (SSPs), with observations, that explicitly resolve single stars within stellar populations.
Our procedure consists in populating star particles with stars, according to input properties that star particles 
consistently inherit from a cosmological simulation. In this way, it is possible to bridge the gap between the mass 
resolution of state-of-the-art cosmological simulations, with star particles that have a typical mass ranging 
between $\sim10^8$ and $\sim10^3$ M$_{\odot}$, and the mass suitable for an appropriate comparison with 
observations of single stars. Features of parent star particles such as mass, position, 
and their age-metallicity distribution are supplied to a code named TRILEGAL, along with an initial mass 
function (IMF). 
TRILEGAL then generates a catalogue of stars that share properties with their parent population and 
that have a mass ranging between $\sim 0.5$ and few M$_{\odot}$. Also, stars are provided with photometric 
properties, so that we know their absolute magnitude and can retrieve their apparent magnitude in different bands. 

We have focused on a possible application of this new approach: we have simulated the observable stellar 
content of a volume that is comparable to the solar neighbourhood of the Milky Way (MW). 
First, we carried out a cosmological 
hydrodynamical simulation of a disc galaxy and selected star particles located within a sphere around the galactic 
plane, $\sim 8$ kpc far from the galaxy centre. This region can be approximately deemed as a representation 
of the solar neighbourhood. We then provided the TRILEGAL code with properties of these parent star 
particles, and it generated a catalogue of mock stars. We conceived a procedure to select generated stars in 
order to match the main observational constraints of one of the survey, namely the Geneva-Copenhagen Survey 
(GCS), that recently collected 
informations of several thousands of nearby stars, deriving accurate determinations of their physical properties. 

The most relevant results of our work can be summarised as follows.
\begin{itemize}
\item We constructed and presented colour-absolute magnitude and colour-apparent magnitude diagrams 
for stars in our sample, showing the observational features that stars originated from selected star particles in a 
cosmological simulation would have.

\item We investigated the age-metallicity distribution for our star sample. 
We find that young and intermediate-age stars 
are characterised, on average, by a higher metal content than old stars. 
Moreover, old stars have a spread in metallicity that is remarkably wider ($\sim 2.5$ dex) than younger 
stars ($\sim 1.5$ dex), and the metallicity distribution progressively narrows towards the youngest ones, 
thus in keeping with indications from observational data. 

\item As for the mass-metallicity distribution, we find that stars tend to move towards smaller masses and lower 
metallicities as the stellar age increases, in agreement with observations.
Also, we successfully generated a wide spectrum of low- and intermediate-mass stars. The bulk of stars 
in our sample have a mass ranging between $0.6$ M$_{\odot}$ and $1.5$ M$_{\odot}$, their mass never being
larger than $7$ M$_{\odot}$. Stellar mass distribution is more clustered as one moves from young to 
intermediate-age to old stars, only few old stars being more massive than $1$ M$_{\odot}$. 

\item We examined the metallicity distribution function (MDF) of stars in three different cosmological simulations 
that we carried out: 
the considered star samples represent the observable stellar content of three solar neighbourhood-like volumes.
We compared our predictions with chemical properties of stars in the solar neighbourhood in the GCS, 
whose MDF has a slightly sub-solar mean value of $[Z/H]=-0.02$ \citep{Casagrande2011}.  
The MDF of the star catalogue of our reference simulation AqC5--fid has a mean $[Z/H]= 0.04$, that 
is in good agreement with observations. 
The means of the two distributions differ by only $0.06$ dex, such difference implying that stars of our sample 
are on average richer in metals than observed ones by $\sim 15 \%$. 
The considered star sample of AqC5--fid is characterised by a 
low-metallicity tail that is more extended than the MDF of the GCS, and that is mainly populated by old stars. 
Our results are in keeping with observations, the peaks of the two MDFs perfectly overlapping and 
pinpointing a solar metallicity.
We also consider the MDFs of synthetic star catalogues in simulations AqC5--cone and AqC5--3sIMF.
Stars in AqC5--cone have a MDF with slightly super-solar median and an excess of metal-rich stars, 
due to the moderate low-redshift star formation rate, that further enriches in metals outer regions of the galaxy disc. 
As for the simulation AqC5--3sIMF, characterised by a Kroupa IMF with a third slope limiting the number 
of massive stars, the peak of the MDF is subsolar (mean $[Z/H]= -0.53$). 
The comparison with the GCS suggests that stars in the solar neighbourhood are metal richer than 
those in AqC5--3sIMF by a factor of $\sim 2.5$. 
A straightforward comparison between predicted and observed MDFs can be however not trivial, 
because of sample selection effects that affect observational data sets.
When slicing our star samples into different subsamples according to stellar age, we find that old stars 
are responsible for the low-metallicity tail of the MDF, in agreement with observations. 

\item Properties of stars that are originated from star particles in 
marginally different regions of the simulated galaxy are remarkably similar. Stars at fixed distance 
from the galaxy centre are characterised by comparable MDFs, despite their exact position angle 
on the galactic plane. Also, we find that stars within volumes centred at marginally different 
distances (i.e. within $1$ kpc, on the galactic plane) from the galaxy centre with respect to the 
reference $r_{\rm Sun}$ share very similr MDFs. 
A natural prediction of this investigation is that chemical properties of stars observed in the solar 
neighbourhood are typical of MW stars at a similar distance from our Galaxy centre. Moreover, chemical 
features of nearby stars are shared to some extent by stars beyond the conventional extension of the solar 
neighbourhood.
\end{itemize}

In the present work we have explored the feasibility of the approach we have developed and we have 
shown its capabilities and possible applications. Its ultimate goal is to introduce a new technique that allows us to
perform a comparison as accurate as possible between simulation output and observations of stars. 
This method is versatile and can be tailored to fit output of different particle based codes. We envisage 
that it will have interesting applications when comparing simulations of galaxy formation, reaching an ever 
increasing resolution, with ongoing and future survey data releases, such as Gaia-ESO and especially GAIA.

Here we would like to highlight another possible future application of this tool. 
Cosmological simulations allow us to identify peculiar classes of stars at different redshifts, during the formation
of a galaxy. For instance, stars belonging to substructures of the galaxy disc or halo, to captured or 
destroyed stellar satellites, or stars being members of stellar streams: these stars likely have chemical features 
and patterns that are different from the ones shared by the bulk of stars in a component of the galaxy, 
since they did not take part to the general galaxy evolutionary path. These stars can be tracked during the simulation. 
In this way, it could be interesting to construct the CMDs of stars that have had a unique life. 
Therefore our approach can have a significant comparative and predictive power.
Also, it could be used to evaluate selection effects and sample completeness of observed catalogues.

Another possible interesting direction of future investigation will be to predict properties of stars 
with distinguishing chemical features and to compare predictions from hydrodynamical models with data: for instance, 
we can study peculiar classes of stars by selecting generated stars according to physical criteria 
(e.g. surface gravity) that different tracers of stellar populations (e.g. AGB stars, horizontal branch stars, dwarf stars, variable stars) meet. Also, we could select star particles according to chemical properties on the age-metallicity 
relation plane, thus identifying parent particles of peculiar stars, and then retrieve information on their evolution and 
past history.

\section*{Acknowledgments}
We thank the anonymous referee for a prompt, careful and constructive report.
We are greatly indebted to Volker Springel for giving us access to the
developer version of the GADGET3 code. 
We thank Gian Luigi Granato for useful discussions and comments, 
and Donatella Romano for kindly providing us with stellar yields.
Simulations were carried out using ULISSE at SISSA and Marconi at CINECA (Italy). CPU time has been 
assigned through projects Sis17\_AGNFeed and Sis18\_bressan under Convenzione SISSA, 
through the project INA17\_C1A00, 
and through Italian Super-Computing Resource Allocation (ISCRA) proposals and an agreement 
with the University of Trieste. 
The post-processing has been performed using the PICO HPC cluster at CINECA through our expression of interest. 
LT has been funded by EU ExaNest {\sl{FET-HPC}} project No 671553.




\bibliographystyle{mnras} 
\bibliography{cool_ref}





\bsp	
\label{lastpage}
\end{document}